\documentclass[a4paper,11pt]{article}
\pdfoutput=1 

\usepackage{jheppub} 

\usepackage[T1]{fontenc} 

\title{\boldmath The area and  volume laws for  entanglement of  scalar fields in flat  and cosmological  spacetimes}

 \author{K. Andrzejewski}
 \affiliation{University of  Lodz, Faculty of Physics and Applied Informatics,\\  Lodz,  Pomorska 149/153, 	90-236, Poland}

\emailAdd{krzysztof.andrzejewski@uni.lodz.pl}

\newcommand{\be}{\begin{equation}}
	\newcommand{\ee}{\end{equation}}

\newcommand{\mB}{{\mathcal B}}
\newcommand{\bx}{{\bf{x}}}

\abstract{We study the  area  and volume laws   for entanglement  of   free quantum  scalar  fields. In addition to   the entropy, we use  the notion of the capacity  of entanglement,  which  measures  entropy fluctuations.  We consider    flat   spacetimes  as well as the curved  ones  relevant for cosmology. Moreover,  we put  special emphasis on   quench  phenomena and different geometries	 of the  entangling surfaces.   
	\par 
	First,  we show that, in the Minkowski spacetime,   the  capacity of entanglement, like entropy, exhibits the area law for two kinds of geometries of the   entangling surfaces:  the sphere and strip. Moreover, we show that   the ratio of both quantities takes  the  same values  for both  surfaces. 
	Next, we turn our attention to quenches. Namely, we analyse the dynamics of capacity; in particular,   contribution of the volume and surface terms. Moreover, we compare these results with theoretical predictions resulting  from the quasiparticles model.  
	In the second part, we consider the above issues  for the   FLRW  spaces; especially, for    de Sitter space as well as a  metric   modeling the    transition to radiation-dominated era.   
	Finally, we analyse the abrupt  quenches  in de Sitter space.}

\begin{document} 
\maketitle
\flushbottom

\section{Introduction}
\label{s0}
Quantum entanglement is one of the fascinating aspects of quantum systems, and its implications   go beyond quantum information processing and technology.    For example, it seems to play the relevant role  in the  statistical mechanics, condensed matter  or high energy physics, see \cite{b1a}-\cite{b1e} for  wider   discussions and further  references.    The latter aspect is particularly interesting because of quantum gravitational effects. As a result,  the notion of entanglement in quantum field theory has been extensively investigated in the  recent years (especially  for the  low-dimensional or flat spacetimes).    An important example of such studies is the concept of entropy in black hole physics; in particular, its relation to  the  area   instead of  volume scaling, see the pioneering  works    \cite{b4a,b4b,b4c} and  \cite{b3a,b3b} for review, or more generally the notion   of the  holographic entanglement  entropy   \cite{b5a,b5b}. 
\par 
To  quantify  entanglement several measures  have been proposed.
The most popular one, for a pure state of a bipartite system,     is   the von Neumann  entropy.      One of the  interesting properties of the von Neumann entropy   is that  it  can be written  as the expectation value  of  the so-called modular Hamiltonian  (i.e. the minus logarithm of the density operator).      In view of this, the  variance of the modular Hamiltonian    can serve as a measure of   fluctuations of  the entanglement entropy  \cite{b7}. On the other hand,  it turns out that  this variance can be  treated  as a kind of  ``heat" capacity.   
Such  an approach   originally  appeared in the context of  condensed mater physics \cite{b8}.   However,    recently  such  a (modular) capacity    gained  some additional  attention due to     quantum gravitational effects,    holographic   duality, and other aspects of  the field theory, see e.g.    \cite{b6a}-\cite{b6o}.
\par 
Motivated by these results,  in this work we continue the study of the  entropy and capacity of  entanglement, however,   with the  emphasis  on  higher dimensional (in particular (1+3)-dimensional)  flat and non-flat spacetimes as well as various geometries of the  entangling surfaces.  Moreover, we put   special attention on the dynamics of these quantities during quenches; in particular, the analysis of the  area and volume laws. Such  investigations  are relevant for  various physical processes, including   thermalization or non-equilibrium systems, see e.g.  \cite{b12a}-\cite{b12g}.  
\par
The work  is organized as follows.  In Sec.   \ref{s1}  we recall the discretization procedure for  fields in non-stationary  spacetimes  as well as  the  formalism needded  to compute the entropy and  capacity in this approach.   In Sec. \ref{s2} we investigate the  capacity for $(1+3)$ and $(1+2)$-dimensional Minkowski spacetimes  and the spherical entangling surface; we consider  the case of   constant mass  as well as  mass quenches. In the latter case we compare the numerical results with the quasiparticles model  (basing on EPR pairs).
To analyse  universal features of the quasiparticles model as well as  of the ratio of the  capacity and entropy in Sec. \ref{s3} we make analogous considerateness  for the strip  geometry. 
In Sec. \ref{s4}  we  consider  curved spacetimes. We focus on the  FLRW metrics  because of  their cosmological applications and holographic aspects. In particular, we  investigate the above issues for    de Sitter (dS) space as well as after  transition to the  radiation-dominated era. Finally, we consider quenches in  dS space. The conclusions are collected in Sec. \ref{s5},   while  some technical details  and  auxiliary facts are  provided   in Appendices \ref{s6} and \ref{s7}. 

\section{Setup} 
\label{s1}
Let us consider the free scalar field $\Phi$ in the curved  spacetime $g_{\mu\nu}$,  described by  the action 
\be
\label{ac}
-\frac 1 2 \int d^4x\sqrt{-g} (g^{\mu\nu}\partial_\mu\Phi\partial _\nu\Phi +m^2(x^0)\Phi ^2),
\ee
where $m(x^0)$ is, in general, a ``time-dependent" mass parameter   modeling    the quench. Obviously,  the simplest and most relevant case  is the Minkowski   spacetime.  However,  other spacetimes  can be also considered; for example, the FLRW metrics are of great interest,   due to  their applications in cosmology,     or  more generally  time-dependent spherically symmetric spacetimes.  For such   spaces  we can perform   the discretization  procedure  of the action \eqref{ac}, see e.g. \cite{b13a,b13b,b13c}.  As a result   we get  the  Hamiltonian on the lattice\footnote{Throughout  our considerations we put  the lattice spacing equal one.} (together with  appropriate  boundary conditions).  Namely,  in $1+1$ dimensions the resulting Hamiltonian  can be written in  the form   
\be
\label{hd}
H(x_0)=\frac 1 2\sum_{j=1}^{N}\pi_j^2+\frac 1 2\phi^T \Lambda (x_0)\phi,
\ee
where  $\phi =(\phi_1,\ldots,\phi_N)$ and $\Lambda(x_0)$ is a  symmetric $N\times N$  matrix  build with suitable frequencies and couplings parameters; in  higher dimensions the Hamiltonian  is  the  sum  of the ones given in  eq. \eqref{hd}, see e.g. \cite{b13a}.
In view of the above, the discretization procedure  enables to  analyse  various  aspects of the  field theory  and even for  constant mass  leads to  interesting issues.  One of them is the  meaning  of entanglement in  quantum field theory.  In this case,   we   divide the space in two regions.        In the lattice approach, this corresponds to  a
splinting of  the whole system into two parts (consisting with  $n$ and $N-n$ oscillators, respectively). Next,  we   define  the reduced density operator with respect to the  one part.
\par  To quantify quantum correlations
various measures have been proposed.  The von Neumann  entropy, or more generally, the R\'enyi entropy $R_\alpha$ are the most common. It is worth noting  that   the von Neumann  entropy   can be written  as the expectation value of the operator $K=-\ln(\rho)$, the so-called modular Hamiltonian.   In view of this the  variance of $K$, i.e.     $C=\langle K^2\rangle -\langle K\rangle ^2$, can be considered as a measure of the  fluctuations of the entanglement entropy. On the hand, following a  thermodynamical analogy, see e.g.  Refs. \cite{b8,b6a,boer},  $C$ can be treated as  the ``heat" capacity; this in turn   leads to  an equivalent definition and terminology for $C$, namely the   (modular) capacity: 
\be
\label{ec}
C=(\partial_\alpha^2((1-\alpha)R_\alpha))|_{\alpha=1}\ .
\ee   
\par
In this work we will study   the entanglement   entropy and its fluctuation, i.e.  capacity of entanglement,  for (quenched) fields  in some (curved) spacetimes.  In view of the previous  discussion concerning   the lattice procedure these  problems can be  reduced  to  the ones for the discretized systems. Thus, first, we  briefly  recall   the main steps  of such an  approach \cite{b6l,b10a,b15}. Namely,  we  start  with the  instantaneous ground state (at  some initial time, subscript $0$ in the notation) of the whole system. Then, the evolution of the  density matrix\footnote{ For simplicity of notation  we omit    the  time parameter $t$  in the matrices and subsequent quantities and dot refers to derivative with respect to $t$.}    is given by     
\be
\rho(\phi,\phi')=\sqrt{\det(\Omega/\pi)}\exp( i\phi^TB \phi-i{\phi'}^TB\phi'- \frac 12 \phi^T\Omega \phi- \frac 12 {\phi'}^T\Omega \phi'),
\ee
$\Omega=U^T\sqrt{\tilde\Lambda}U$, $B=U^T\tilde BU$  where $\tilde B,\tilde \Lambda$  are diagonal matrices  with elements   $(\tilde\Lambda)_{ij}=\lambda_i^0/b_i^4\delta_{ij} $ and $(\tilde B)_{ij}=\dot b_i/(2b_i)\delta_{ij}$, respectively,  while   $b_j$ are the  solutions  of the Ermakov equations with the frequencies $\lambda_ j$ 
\be
\label{ee9}
\overset{..}{b_j}+\lambda_jb_j=\frac{\lambda_j^0}{b^3_j}, \quad j=1,\ldots,N;
\ee
and, finally,  $U$ is a time-independent matrix   diagonalizing $\Lambda$, i.e. $ U\Lambda U^T=Diag(\lambda_1,\ldots, \lambda_N).$ 
Next, we  split  the whole system into two parts: the first one  $\mathcal{A}$  consisting  of the first $n$  oscillators  and  the second one $\mathcal{B}$  related to the remaining  $N-n$ ones. To  find the reduced density we  rewrite $\Omega$ and $B$ in the form
\be
\Omega=
\left(
\begin{array}{cc}
	\Omega_1&\Omega_2\\
	\Omega_2^T&\Omega_3
\end{array}
\right),\quad B=
\left(
\begin{array}{cc}
	B_1&B_2\\
	B_2^T&B_3
\end{array}
\right),
\ee
where $\Omega_1,B_1$ are $n\times n$ matrices. Next, integrating over the subsystem $\mathcal{A}$ we get  the reduced density of the subsystem $\mB$ 
\be
\rho_{\mathcal {B}}(\phi_{\mathcal B},\phi_{\mathcal B}')= A\exp(i\phi_{\mathcal B}^TZ\phi_{\mathcal B}-i{\phi_{\mathcal B}'}^TZ\phi_{\mathcal B}'-\frac 12 \phi_{\mathcal B}^T\Upsilon \phi_{\mathcal B} -\frac 12{\phi_{\mathcal B}'}^T\Upsilon \phi_{\mathcal B}' +\phi_{\mathcal B}^T\Delta\phi_{\mathcal B}'),
\ee
where $\phi_{\mathcal B}=(\phi_{n+1}\ldots,\phi_N)$ and  $Z,\Upsilon ,\Delta$ are $(N-n)\times (N-n)$ matrices  given by 
\begin{align}
	Z&=B_3-B_2^T\Omega_1^{-1}\Omega_2,\\
	\Upsilon &=\Omega_3-\frac 12 \Omega_2^T\Omega_1^{-1}\Omega_2+2B_2^T\Omega_{1}^{-1}B_2,\\
	\label{ee20a}
	\Delta&= \frac 1 2 \Omega_2^T\Omega_1^{-1} \Omega_2+2B_2^T\Omega_1^{-1}B_2+i\Theta,
\end{align}
with $\Theta=\Omega_2^T\Omega_1^{-1}B_2-B_2^T\Omega_1^{-1}\Omega_2$.  The  spectrum of the  density  operator  with the Hermitian  matrix $\Delta$   was  discussed,   in Ref. \cite{b15}.  
It turns out that it     is of the form 
\be
\label{ee20}
(1-\xi_1)(1-\xi_2)\ldots(1-\xi_{N-n})\xi_1^{m_1}\xi_2^{m_2}\ldots \xi_{N-n}^{m_{N-n}},
\ee
where $\xi$'s are the inverse of the  eigenvalues  (larger than one)  of the following matrix
\be
\label{ee21a}
\left(
\begin{array}{cc}
	2\tilde \Delta^{-1}&-\tilde \Delta^{-1}\tilde \Delta^T\\
	I & 0
\end{array}
\right),
\ee
where 
\be
\label{ee21}
\tilde \Delta=(\tilde\Upsilon)^{-1/2}\Pi \Delta \Pi^T  (\tilde\Upsilon)^{-1/2},
\ee
while $\Pi$ is  an orthogonal matrix   diagonalizing  $\Upsilon$, i.e.  $\Pi\Upsilon\Pi^T=\tilde \Upsilon$.  Now,  from  \eqref{ee20}, definition of the von Neumann  entropy and  eq.    \eqref{ec} we  get   
\be
\label{eee7}
S=-\sum_{j=n+1}^N\left(\ln(1-\xi_j)+\frac{\xi_j}{1-\xi_j}\ln(\xi_j)\right),    \quad  C=\sum_{j=n+1}^N  \frac{\xi_j\ln^2(\xi_j)}{(1-\xi_j)^2}.
\ee 
As noted above, in  higher dimensional spacetimes the total  Hamiltonian is a sum  of independent Hamiltonians of the form \eqref{hd} (see, e.g.,  \cite{b13a}), thus entropy and capacity are also  the sum of the corresponding components.  For example, in $1+3$ dimensions  the discretization procedure  yields $S=\sum_{lm}S^{lm}$  and  $C=\sum_{lm}C^{lm}$   where $S^{lm}$ and  $C^{lm}$  are of the  form as above, see Appendix \ref{s6} for more details.    
\par  
Alternatively, to obtain  the entropy and capacity,   we can  use another approach   based on the correlation (covariance) matrix and symplectic  spectrum. This approach seems more useful for numerical computations and has been successfully applied in various investigations, more details  can be found, e.g.,  in Refs. \cite{cotler,b12g}. Therefore,   we also apply  this approach to numerical calculations.  However,  to obtain some analytical results we will employ    the aforementioned   approach  based on  the eigenvalues of the reduced density. Finally, let us note that     the relation between both the approaches   has been recently discussed  in more detail in Ref. \cite{b16}.         
\section{Minkowski spacetime - spherical geometry} 
\label{s2}
\subsection{$(1+3)$ dimensions }
\label{s2a}
\begin{figure}[!ht]
	\begin{center}
		\includegraphics[width=0.99\columnwidth]{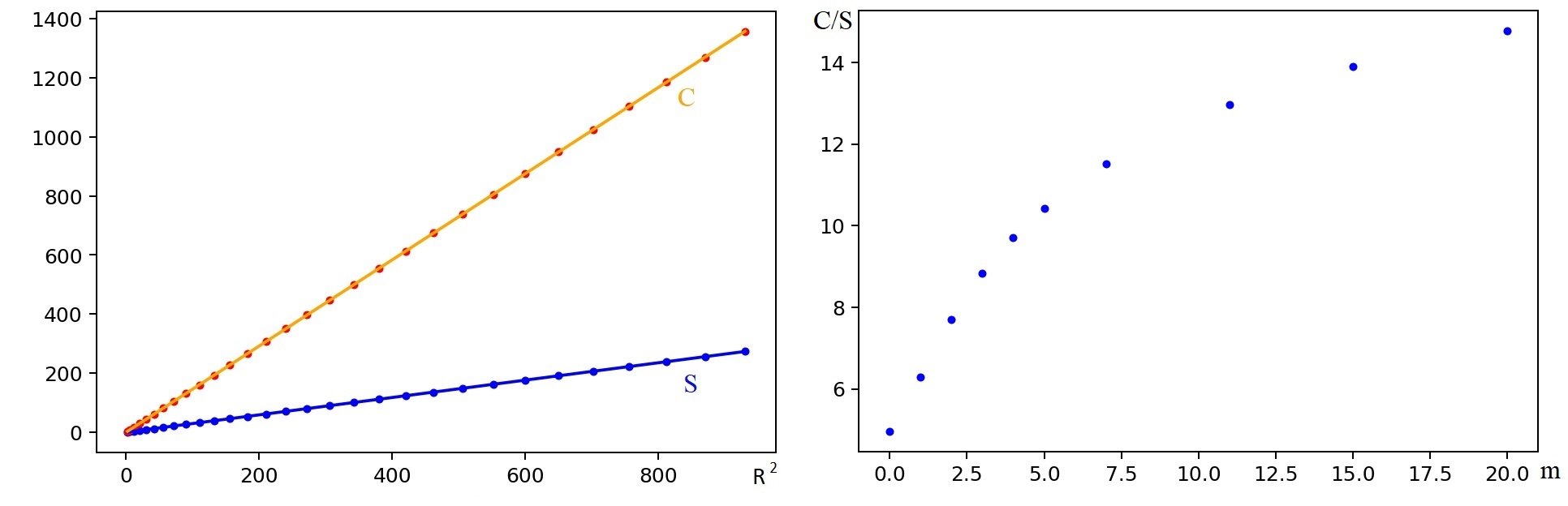}
	\end{center}
	\caption{\small{The $(1+3)$-dimensional Minkowski spacetime - the spherical geometry. The left panel: $m=0$,  entropy (blue data points) and capacity (orange data points) with respect to $R^2$. The right  panel: the ratio $C/S=a_2/a_1$ with respect to $m$, for $m=0$ it is $5.2$.  
			\label{f1}}
	}
\end{figure}

A remarkable   property  of  the  (geometric) entanglement entropy  of the scalar field with   constant  mass (at its   ground state) is that    it obeys the area law,  i.e. at the leading order  the  entropy    is proportional to the area of the   entangling sphere   $S=a_1R^2$ \cite{b4a,b4b,b4c}; in the notation of  Sec. \ref{s1}, 
$R=n+1/2$. In turns out that the same situation  holds for the  capacity.  In fact,  plotting the   capacity\footnote{In our considerations  we take $l$'s  so large  that the entropy and capacity do not change significantly, i.e. a few thousand; this is consistent   with other   considerations reported in the literature, see e.g. \cite{b6n} and references therein.   }  as the function of $R^2$  for the  ground state of the  discretized Hamiltonian, see Sec. \ref{s1}  and Appendix \ref{s6},    we observe, in  the  left part of  Fig. \ref{f1},  that   $C=a_2R^2$.    However, let us stress that  the slopes $a_1,a_2$  for the  entropy and capacity, respectively,  are different and both depend on mass.    In view of this   the ratio  $C/S$, at the leading order, does not depend on $R$ ($C/S=a_2/a_1$). Now, let us consider  this ratio  for various values of the mass parameter, see the right panel in Fig. \ref{f1}. For the massless field the slope of   capacity is $a_2\simeq 1.56$  while  for  entropy $a_1\simeq 0.3$, thus the ratio $C/S$  is approximately equal to $5.2$  what  coincides with the results of Ref. \cite{boer}; next, it increases with mass, see Fig. \ref{f1}.    In general, this ratio is scheme dependent; though, for  some conformal  theories (with a dual  holographic  gravity description)   is  universal and  equal to one, see Ref. \cite{boer} for a more extensive discussion on this subject.  Here, we  will  analyse another aspect of $C/S$. Namely, the dependence on the  geometry of the entangling  surface. To this end  in Sec. \ref{s3} we will make a similar analysis for  the strip geometry  and show that  both cases  coincide very well; this, in turn,  suggests  another  universal property of this ratio.
\par 

\begin{figure}[!ht]
	\begin{center}
		\includegraphics[width=0.99\columnwidth]{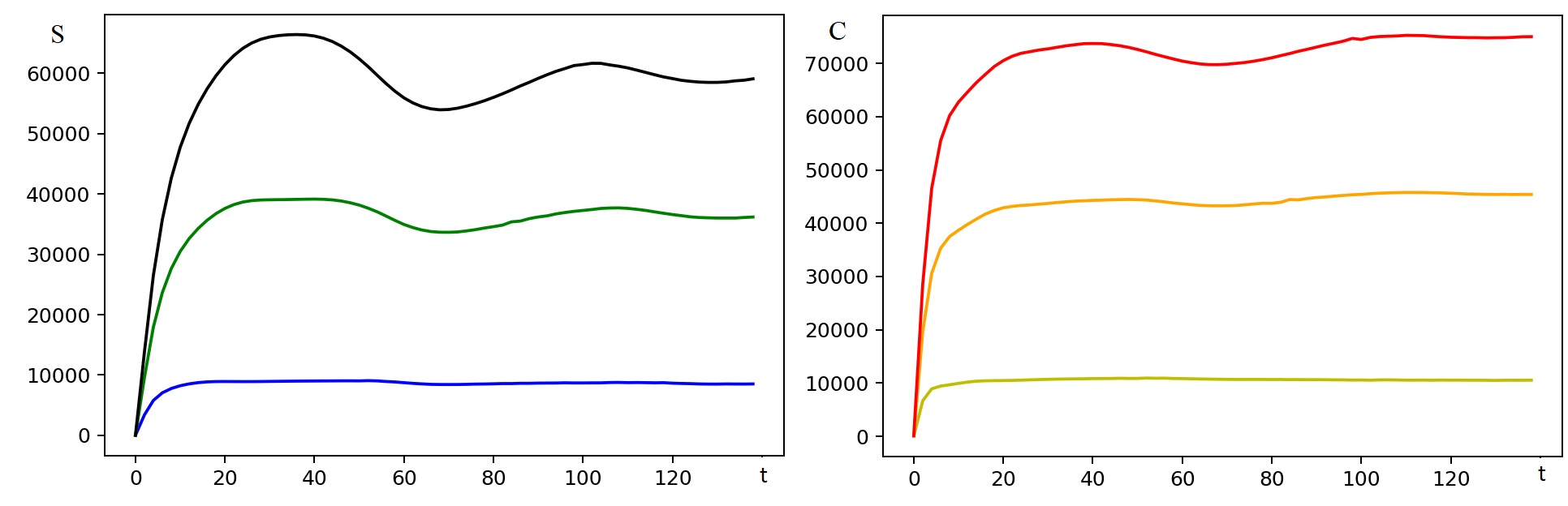}
	\end{center}
	\caption{\small{ The abrupt  quench in  the $(1+3)$-dimensional Minkowski space - spherical geometry;   $m_i=10$, $m_f=0$, $N=60$.     The left panel:  entropy  - blue $n=15$, green $n=25$, black $n=30$. The right panel: capacity  -  yellow $n=15$, orange  $n=25$, red $n=30$.    
			\label{f2}}
	}
\end{figure}

\begin{figure}[!ht]
	\begin{center}
		\includegraphics[width=0.99\columnwidth]{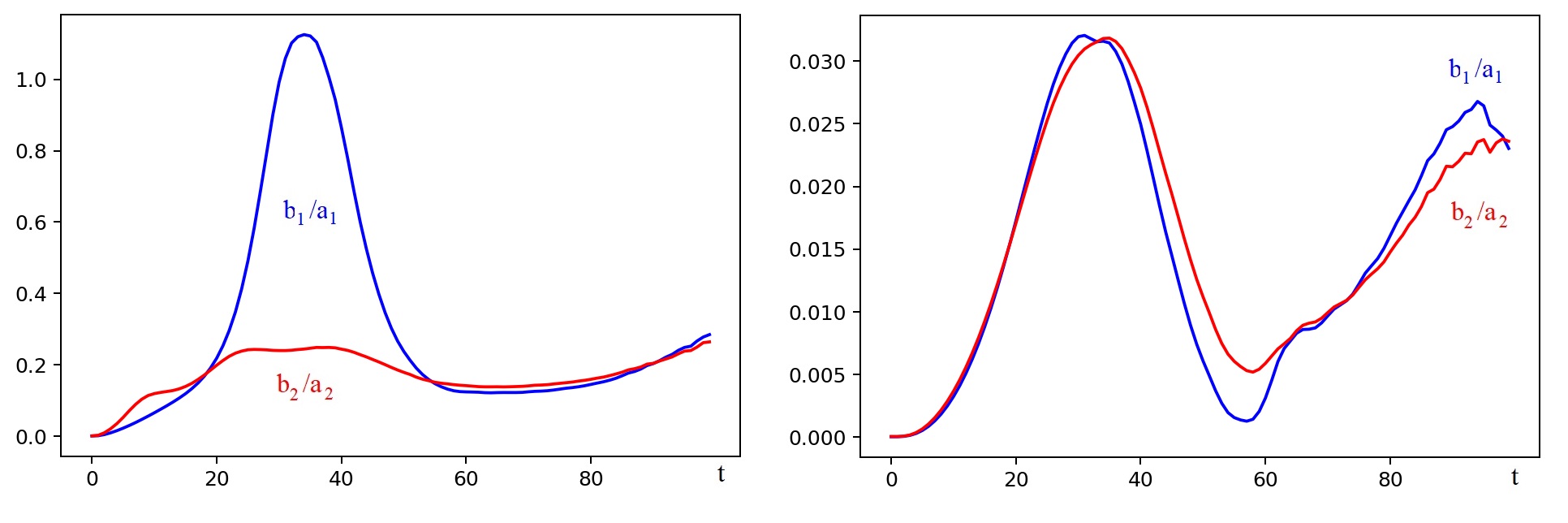}
	\end{center}
	\caption{\small{The ratio   $b_i/a_i$  for the abrupt  quench in the   $(1+3)$-dimensional Minkowski space - spherical geometry;   $N=60$,   $m_f=0$.  Blue curve - $b_1/a_1$ for   entropy, red curve -  $b_2/a_2$  for capacity.   The left panel:   $m_i=10$. The right panel: $m_i=0.5$. 
			\label{f2a}}
	}
\end{figure}

\begin{figure}[!ht]
	\begin{center}
		\includegraphics[width=0.99\columnwidth]{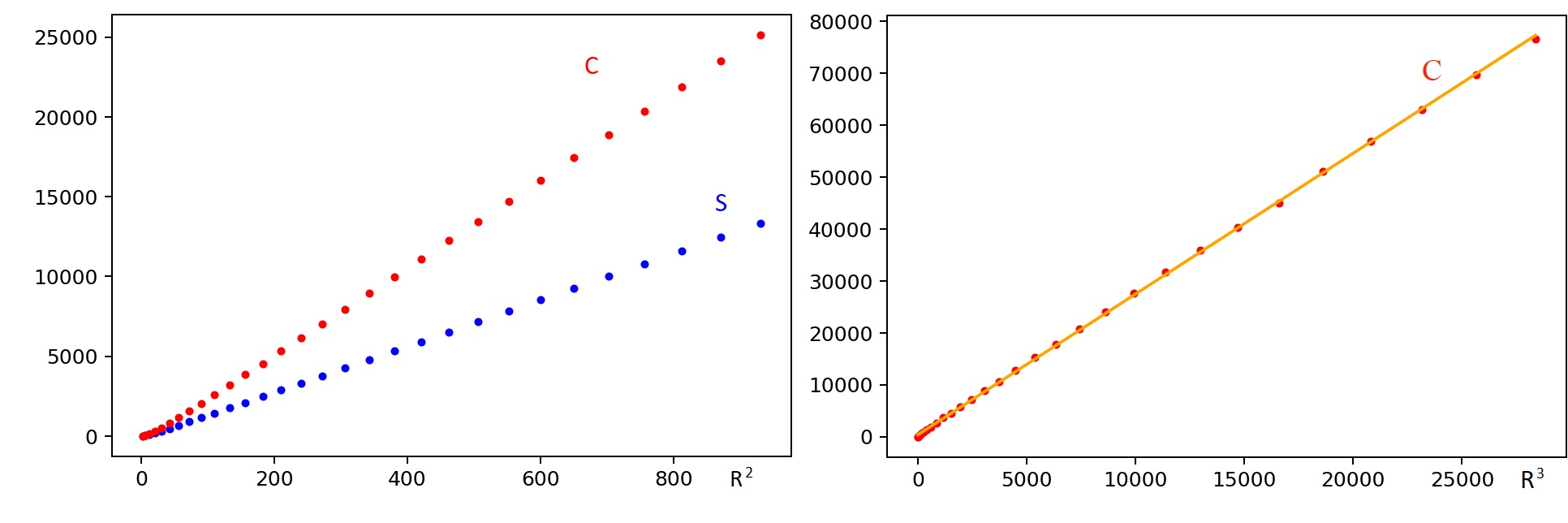}
	\end{center}
	\caption{\small{The abrupt  quench in  the $(1+3)$-dimensional Minkowski space - spherical geometry; $N=60$, $m_f=0$.  The left panel: $m_i=10$,    $t=2$ slice;   entropy (blue points)  and capacity (red points)   with respect to the area,  $R^2$.    The right  panel: $m_i=10$, $t=30$ slice,  the  capacity with respect to the volume, $R^3$.     
			\label{f2b}}
	}
\end{figure}

Now, let us consider a more complicated  situation when mass of the field is   time-dependent.       We will  analyse the abrupt  quench; though the  continuous protocols (e.g. related to hyperbolic tangent) can be also considered.  Moreover, we will focus on the case where  the final mass  is equal to zero (the critical protocol).   So,  we start with a field of  mass   $m_i$   and  next  at time $t_0=0 $   there is a sudden change of mass    to   zero. In this case,  the solutions of the Ermakov equations are given by the formulae 
\be
\label{e16}
b_j(t)=\sqrt{r_j\cos(2t\sqrt{\lambda_j(f)})+s_j},
\ee 
where $r_j=(\lambda_j(f)-\lambda_j(i))/(2\lambda_j(f))$,    $s_j=(\lambda_j(f)+\lambda_j(i))/(2\lambda_j(f))$,  and $\lambda_j(i), \lambda_j(f)$   are the eigenvalues of $\Lambda$  before and after the quench, respectively.    
We begin  our analysis  with the   temporal evolution of the capacity   for various $R$ (equivalently $n$).  The typical dynamics of   the entropy and  capacity  is presented in Fig. \ref{f2}. Let us note that  their values grow with  $n$; namely,  for $n<N/2$ they are smaller  than   for  $n=N/2$. Moreover,  we  see that the    dynamics of the entropy and capacity   have  an  increasing period, up to $t=R=(n+1/2)$, and  finally   they oscillate  around some asymptotic values (depending on $n$).
To analyse the area law, let us make, for   fixed time $t$, the decomposition
\be
\label{dec}
S(t)=a_1(t)R^2+b_1(t)R^3,\quad  C(t)=a_2(t)R^2+b_2(t)R^3.
\ee
Then the coefficients $a_i(t)$ and $ b_i(t)$, for $i=1,2$,   describe the impact of the surface and  volume  terms, respectively. In particular, when   $Rb_i(t)/a_i(t)\ll 1$ the are law holds, in the other case the volume term  has  to be taken into account.     To  analyse  the dynamics of the area law  we plot the ratio $b_i/a_i$ ($i=1$  - entropy, $i=2$ - capacity), see Fig. \ref{f2a}. 
Obviously, the  area law holds  before the quench  (in our case for  $t\leq 0$). However,   for sufficient small times the surface term is also dominant; in fact,  for  $t=2$ and even for  relatively large gap, i.e. $m_i=10$,   the values of the   entropy and capacity with respect to $R^2$ fit quite well to the straight lines, see the left panel of Fig. \ref{f2b}.
Now, let us focus  on  large times. Then the situation depends on the initial mass. When  the gap is large (e.g. $m_i=10$)  then the volume term remains decisive;	   for  smaller gap the mixture of the volume and surface terms is necessary. Indeed,  for  a large initial mass the ratio $b_i(t)/a_i(t)$ oscillate about  some asymptotic values, see also Fig. \ref{f4} in Sec. \ref{s2b}. Finally,  let us not that  the volume term is particularly  crucial  for time $t=N/2=30$,  see the right panel  in Fig. \ref{f2b} where  $m_i=10$, $t=30$  and the plot is with respect to the volume $R^3$. 

\subsection{$(1+2)$ dimensions} 
\label{s2b}
\begin{figure}[!ht]
	\begin{center}
		\includegraphics[width=0.99\columnwidth]{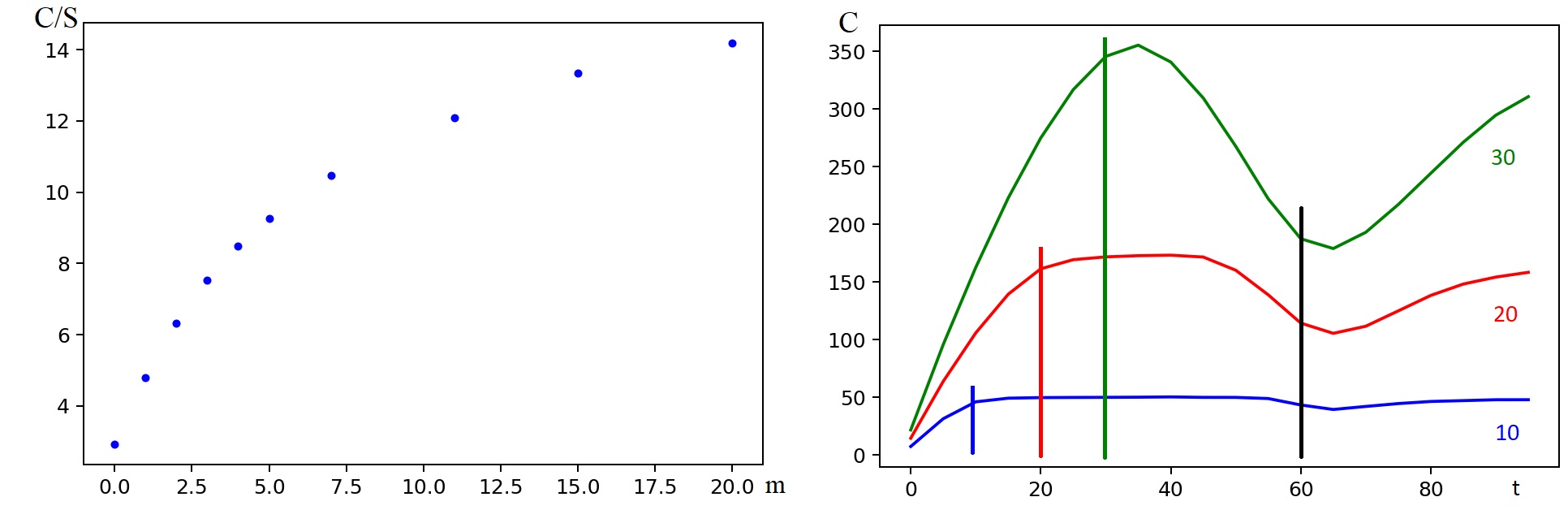}
	\end{center}
	\caption{\small{ The $(1+2)$-dimensional Minkowski space - the spherical geometry. The left panel: the ratio $C/S=a_2/a_1$ with respect to  $m$; for $m=0$ it equals $2.92$. The right panel: the dynamics of the capacity for  the abrupt  quench,   $m_i=10$, $m_f=0$,  $N=60$;  			      blue $n=10$, red $n=20$, green $n=30$ (vertical lines correspond to  periods:  $t=n$). The black vertical line corresponds to $t=N=60$ -  a partial revival.     
			\label{f3}}
	}
\end{figure}

For two  spatial dimensions   and constant  mass, the numerical computations yield that   the  capacity like   entropy  linearly increases with the radius $R$ of the sphere (for the ground state of the discretized Hamiltonian), i.e.  $S=a_1R$ and  $C=a_2R$. However, as in three spatial dimensions  the slopes ($a_1$ and $a_2$) for both of them are different and depend on mass. Using  results from Sec. \ref{s1} and Appendix \ref{s6}, we  plot the ratio $C/S$ (at the leading order, this is equal to  $a_2/a_1$) with respect to $m$, see the left panel in  Fig. \ref{f3}. For the massless case this ratio   is equal to $2.92$ and then increases with  mass.  Roughly,  the values of this ratio are  smaller than in the $(1+3)$-dimensional case, cf. Fig. \ref{f1}, though for a larger $m$ the difference becomes negligible. We will return to this relation in Sec. \ref{s3} where the strip geometry in $1+2$ dimensions  will be analysed.   
\par
\begin{figure}[!ht]
	\begin{center}
		\includegraphics[width=0.99\columnwidth]{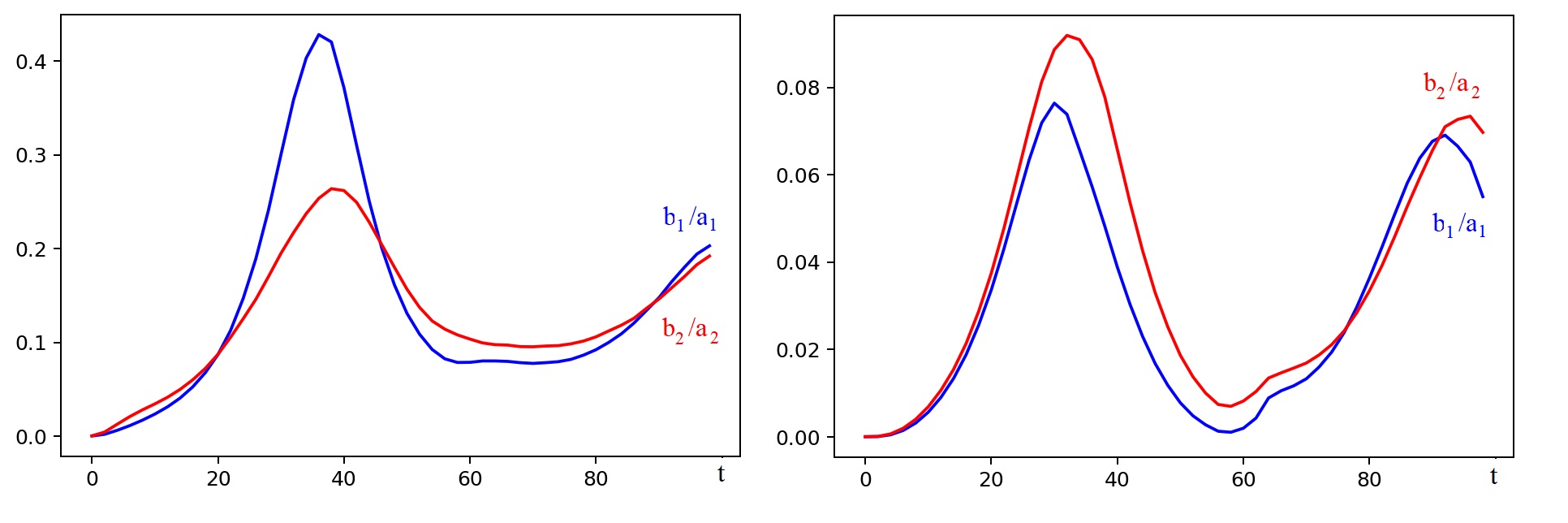}
	\end{center}
	\caption{\small{The ratio   $b_i/a_i$   for the abrupt  quench in  the $(1+2)$-dimensional Minkowski space - spherical geometry; $N=60$,   $m_f=0$.  Blue curve - $b_1/a_1$ for  entropy. Red curve - $b_2/a_2$ for  capacity.   The left panel:   $m_i=10$. The right panel: $m_i=0.5$. 
			\label{f3b}}
	}
\end{figure}

\begin{figure}[!ht]
	\begin{center}
		\includegraphics[width=0.99\columnwidth]{
			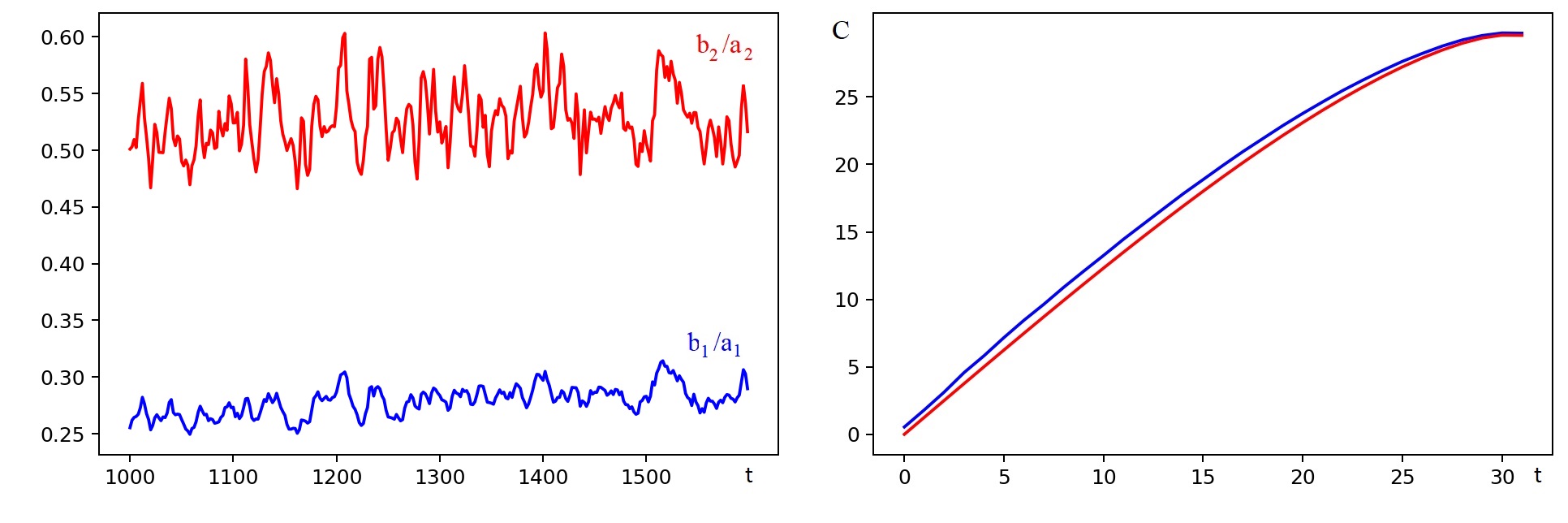}
	\end{center}
	\caption{\small{The $(1+2)$-dimensional Minkowski space - the spherical geometry. Left panel:  the abrupt  quench, $m_i=10$, $m_f=0$,  $N=60$, $n=30$;  the  blue curve - $b_1/a_1$ for  entropy, the   red curve - $b_2/a_2$ for  capacity.   The right panel: 		  evolution of the capacity  for the abrupt quench, $m_i=0.25$, $m_f=0 $,  $N=60$, $n=30$.   The red line - theoretical values based on eqs. \eqref{r12} and \eqref{cc}, the blue line  the numerical results.   	\label{f4}}
	}
\end{figure}

   
Now, let us consider the abrupt quenches. Similarly to the three-dimensional case entropy and capacity increase up to time $t=n$ and then the oscillations begin, see the right panel in Fig. \ref{f3}. More precisely, after the initial growth,    there is  a ``plateau" (with  a very  slow increase) which terminates about  $t\simeq N-n$ and then, after $t\simeq N$,  we have a revival of the entropy and capacity (see the  black vertical  line in Fig. \ref{f3}).    For further times the oscillations are  around some asymptotic  values.  Such a behaviour    can be  interpreted  in terms of the quasiparticles model   presented  in Ref. \cite{b5c}  for the  $(1+1)$-dimensional case and finite-size integrable systems; we will analyse this  issue   in Sec. \ref{s3}. 
Now, we will study the area law.    Namely, using  the decomposition analogous   to the formula \eqref{dec} (in the present case it  contains  $R$ and $R^2$ terms, respectively)  we    plot the ratio  $b_i(t)/a_i(t)$ of the surface and volume terms in the $(1+2)$-dimensional case,  see Fig. \ref{f3b}. Then, for sufficiently short time $t\ll R$ we observe the area law.      
On the other hand, for larger  times the  area law does not hold.     Namely, let us take $m_i=10$ as in   the previous case. Then,  from    the left panel of Fig. \ref{f4}    we infer  that for  a large initial mass the ratio $b_i(t)/a_i(t)$ oscillates, and  the oscillations asymptotically settle down to  some  relatively large  values (in our case  about $0.5$ for the  capacity and $0.2$ for entropy).  Thus,  for large times, the  volume term  is relevant.  This can be also confirmed by  plotting  time slices for   entropy and capacity  as a function of  $R^2$;  in such a case   we obtain quasi-linear behaviour. 

\subsection{Theoretical predictions}
Basing on the quasiparticles (EPR pairs)  model, some theoretical results concerning  the (R\'enyi) entropy  for the abrupt quenches have been obtained in Refs. \cite{cotler,casini}.  We  will use these results to find  a theoretical description of the  dynamics of the  capacity, and next  compare them  with the numerical computations.   Namely, following the considerations of Ref. \cite{cotler}, we obtain that the      R\'enyi entropy in $1+2$ dimensions  takes  the form
\be
\label{r12}
R_{\alpha}=
s_\alpha\left\{
\begin{array}{cc}
	2(t\sqrt{R^2-t^2}+R^2\arcsin(t/R))& t<R,\\
	\pi R^2& t>R;
\end{array}
\right.
\ee
where
\be
\label{r12b}
s_\alpha=\frac{2\gamma_E+\psi(1-1/2\alpha)+\psi(1+1/2\alpha)+2\alpha(\ln 4-1)}{16\pi(\alpha -1)}m^2.
\ee
while $\psi$ denotes digamma function.  
For $1+3$ dimensions we have
\be
\label{r13}
R_{\alpha}=
s_\alpha\left\{
\begin{array}{cc}
	2\pi(R^2t-t^3/3) &t<R,\\ 
	\frac{4\pi}{3} R^3& t>R;
\end{array}
\right.
\ee
where
\be
\label{r13b}
s_\alpha=\frac{4\alpha-3\cot(\pi/4\alpha)+\cot(3\pi/4\alpha)}{48\pi(\alpha -1)}m^3\, .
\ee 
Taking the limit $s=\lim_{\alpha\rightarrow 1}s_\alpha$ in eqs. \eqref{r12b} and \eqref{r13b}  one gets  $s=m^2\ln(2)/4\pi$ (in two spatial dimensions)	and $s=m^3/(12\pi)$  (in three spatial dimensions), see Ref. \cite{cotler}.  Now, using   formula  \eqref{ec} we  get that  the capacity $C$ has also  the  form of eqs. \eqref{r12} and \eqref{r13}, respectively with the factor $s$   replaced by the constant $c$: 
\be
\label{cc}
c=\frac{\pi m^3}{16}, \quad (1+3)\textrm{ dimensions};\quad\quad  c=\frac{7\zeta  (3)m^2}{16\pi}\simeq 0.167m^2,\quad (1+2)\textrm{ dimensions.}
\ee
Since the above formulae  rely on a relatively simple model,  they  validity is limited and involves several assumptions, see Ref. \cite{cotler}  for more detailed discussion; in particular, the initial mass should be sufficiently  small. Let us compare these models  with our  numerical results.  First, according  to  the considerations from previous sections,  we have  the growth of capacity up to $t\simeq R$.   More precisely, in $d=2$ dimensions  the theoretical  (see eq. \eqref{r12} and below) and the numerical results are   quite consistent when the initial mass is less than  one. This can be seen especially  for higher $n$; namely,  taking $n=N/2$  and initial mass  $m_i=0.5$ we see  in the right panel of Fig. \ref{f4}  that the theoretical and numerical results coincide quite well (the  numerical plots are shifted to agree at $t= R$ ). Moreover, for initial times $t\ll R$  they yield the area law, what    coincide very well  with the numerical results presented in  Sec.  \ref{s2b}.
For, the $(1+3)$-dimensional case the quasiparticles picture yield also the growth up to $t=R$ what coincides with the considerations obtained in Sec. \ref{s2a}. However, in this case the capacity (entropy) dynamics is not  matched so well with the numerical results; this fact  can be related  to an  additional  contribution of the  logarithmic  divergence to  the area law after quench in  $1+3$ dimensions  \cite{wilczek}.  
\section{Minkowski spacetime - strip geometry  }
\label{s3}
In this section  we study the above issues for another geometry of entangling surface.   This is interesting due to  possible universal properties  as well as  validity of the  quasiparticles model. To this end, 
following the reasoning of Ref. \cite{cotler}, we can find  the form of  the capacity for    the strip of width $l=2R$ in $1+d$ dimensions, when tracing over a $d$ dimensional slab  of width $2R$   (the case  $d=1$  corresponds to the interval, see \cite{b6l}). In such a case  the system factorizes and  the entropy as well as    capacity reduce to the integral of their  one dimensional counterparts, see Appendix \ref{s6}. Then using eq. \eqref{ec} we readily  get 
\be
\label{capnu}
C_{strip}=\frac{A_{\perp}}{2^{d-2}\pi^{(d-1)/2}\Gamma(\frac{d-1}{2})}\int_0^\infty dk k^{d-2} C(R,m^2(t)+k^2),
\ee    
where $A_{\perp}$ is the width of the strip in the perpendicular direction  and $C$ is the capacity for the  one dimensional system  (i.e. for the interval of the length $2R$).
\par 
On the other hand, for  $d=2$ and  the  periodic boundary conditions the quasiparticles model \cite{cotler,casini} implies the following form of the capacity
\be
\label{capd}
C_{strip}^t=\frac{4A_{\perp}c}{\pi}
\left\{
\begin{array}{cc}
	t &t<R,\\ 
	t-\sqrt{t^2-R^2}+ R\arccos(R/t)& t>R,
\end{array}
\right.
\ee  
where $c$ is given by eq. \eqref{cc}. For the Dirichlet boundary conditions we should make the replacement $c \rightarrow c/2$ and  the  change  is at $t=2R$. In what follows we will  compare  the theoretical model \eqref{capd} with the numerical  results  based on  eq. \eqref{capnu} as well as with the   previous ones for spheres.

\par
\begin{figure}[!ht]
	\begin{center}
		\includegraphics[width=0.99\columnwidth]{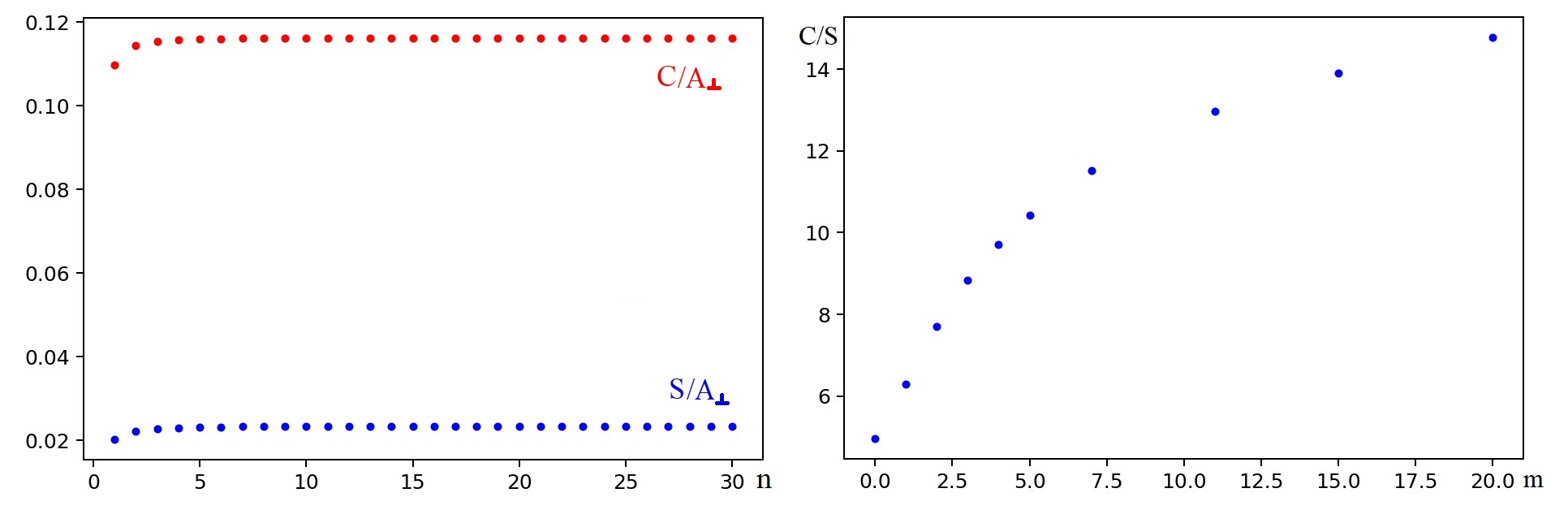}
	\end{center}
	\caption{\small{The $(1+3)$-dimensional Minkowski space - the strip geometry, $N=60$. The left panel $m=0$, entropy - blue points, capacity - red points, with respect to $n$. The right panel: the ratio   $C/S$ (for higher $R$) with respect to $m$,  cf. Fig. \ref{f1}  for the spherical geometry.   
			\label{f6}}
	}
\end{figure}

  
First, let us analyse the case of constant mass.  To this end  we compute the  entropy and capacity with respect to $n$ (equivalently $R$).   To fix attention we present  results   in $d=3$ spatial dimensions, see  the left panel of Fig.  \ref{f6};    for the two spatial dimensions the results are similar.  Namely, we observe that for  sufficient large  radius $R$ both quantities are almost  constant   (here we present the case $m=0$; however, for other values of $m$ this is even more evident) and thus do not depend on $R$.  In consequence, only the transversal area remains and the area law holds.   In view of this the ratio $C/S$ is constant, i.e. it  depends on $m$ only (at the leading order).  Let us compare this ratio with the one obtained for the spherical geometry.  Namely,   in the right panel of  Fig. \ref{f6} we present this ratio as a function of  mass (here  we consider  $(1+3)$-dimensional case,  but a similar situation holds for two spatial dimensions). Comparing the right panels of   Fig. \ref{f1} and Fig. \ref{f6}    we see that  $C/S$    is the same       for both geometries  with very good  accuracy (the same holds  also in   two  spatial dimensions). Thus	despite a quite different geometry of the entangling surface,  $C/S$ does not change. This observation suggests certain universality of this ratio. 
\par
\begin{figure}[!ht]
	\begin{center}
		\includegraphics[width=0.99\columnwidth]{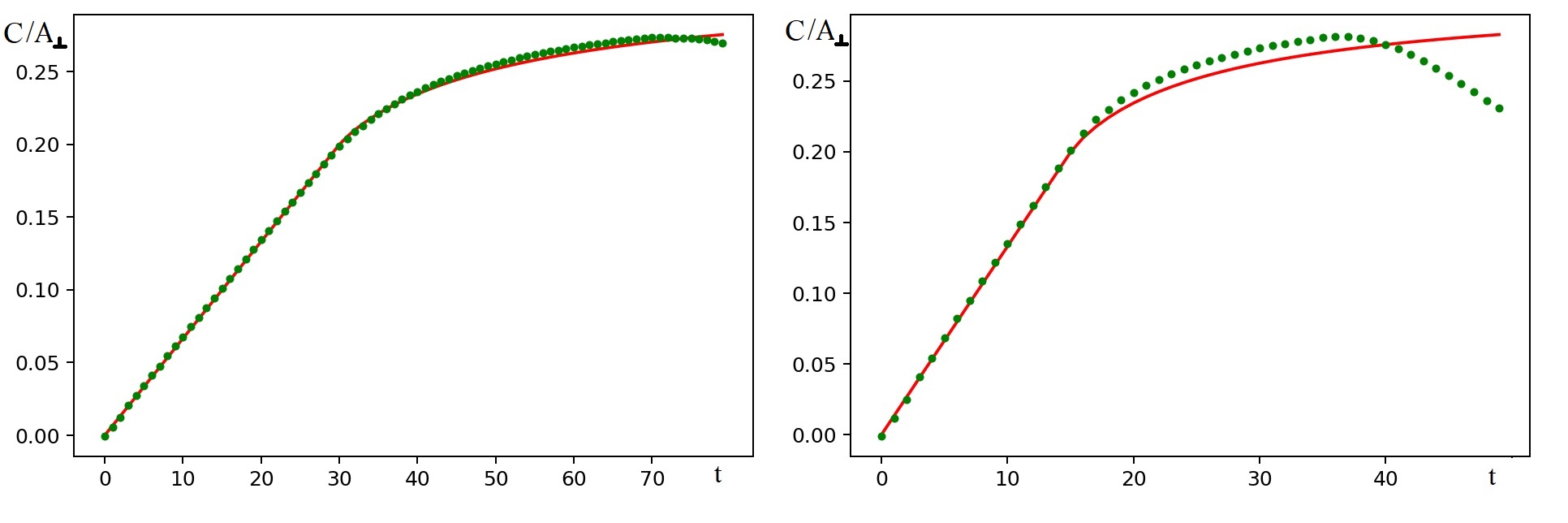}
	\end{center}
	\caption{\small{ The $(1+2)$-dimensional Minkowski space - the strip geometry. The dynamics of the capacity for the abrupt quench $m_i=0.25$, $m_f=0$, $N=100$, $n=30$. The theoretical capacity  -  the red curve, the numerical results -  green points.        The left panel: the Dirichlet boundary conditions. The right panel: the periodic boundary conditions.  
			\label{f8}}
	}
\end{figure}

\begin{figure}[!ht]
	\begin{center}
		\includegraphics[width=0.99\columnwidth]{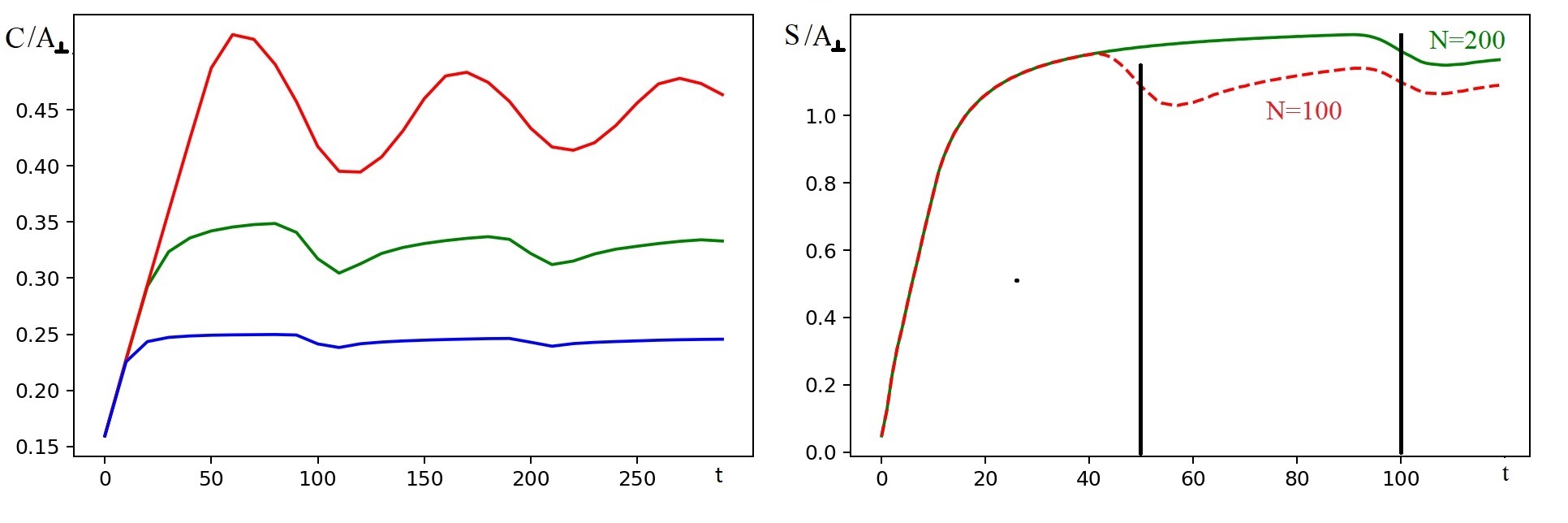}	
	\end{center}
	\caption{\small{The quench in the $(1+2)$-dimensional Minkowski space - the strip geometry. The left panel: the capacity evolution for further times, $m_i=0.25$, $m_f=0$, $N=100$, the Dirichlet boundary conditions. Blue -  $n=10$, green - $n=20$, red - $n=50$.  The right panel: the entropy  dynamics  for the periodic boundary conditions ($m_i=1$ and $m_f=0.01$) for $n=20$;   $N=100$ the red dashed line, $N=200$ the green line.  The black vertical  lines denote  the revival times following from the quasiparticles model (i.e. $n=N/2$).   
			\label{f9}}
	}
\end{figure}

 
Now let us consider  the quenches for the  strip geometry.  Using eq. \eqref{capnu} we can compute the dynamics of the capacity.   In two spatial  dimensions    the numerical results  for the initial  times  are presented, for $n=30$,  by the red curve    in Fig. \ref{f8}.    We observe, in particular, a  linear growth  up to $t=n=2R=30$  for the  Dirichlet (and $t=n/2=15$  periodic, respectively) boundary conditions.  More importantly,   with an appropriate choice of initial mass   (i.e. about $m_i=0.5$)  the slope  agrees  very well with the theoretical predictions following from the formula \eqref{capd},  see the red curve in Fig. \ref{f8} (as usual we shift the plots to match them).        For further time  ($R\ll t $ ), similar to the spherical geometry, the oscillatory behaviour appears, see the left plot in Fig. \ref{f9}. In view of this let us study the role of the volume term.  It turns out that, similarly to the spherical geometry,   for a  larger $m_i$  the contribution related to the  volume factor  becomes more  significant; this can be  especially  seen by considering the time  slices for large times. Namely, taking $m_i=10$  the numerical computations give that  the  capacity increases linearly with $n$ (at least for suitable radius).
For  three dimensional case   we  observe again the linear growth at the initial times; however, the slope  (except small masses) is different  from  the one obtained by means of the quasiparticles model (similarly to  the spherical case). 
\par 
Now, let  us recall  that for  the spherical case, see  Fig. \ref{f3}, we have  some  plateaus and revival times  in the entanglement  dynamics; here, see the left panel in Fig. \ref{f9}, we observe an analogous  situation (for the Dirichlet boundary conditions the first revival time corresponds to   $t=N=100$). To gain some  insight into this issue let us plot the evolution of the entanglement entropy (a similar situation holds for the entanglement  capacity)  for different  values of $N$ and  the periodic  boundary conditions. Namely, in the right panel of Fig. \ref{f9}  we present the cases $N=100$ and $N=200$ with $n=20$. Then, after the linear growth, for  $n/2  \lesssim    t  \lesssim  (N-n)/2$  we observe a plateau with slow saturation. Next, the plateau terminates at $t	\simeq  (N-n)/2$  (the first quasiparticles produced at the boundary of the subsystem re-enter due to periodic boundary conditions). Such a process lasts until  $t\simeq N/2$ when we observe the entanglement revival and the dynamics restarts. In view of this,  the quasiparticles mechanism  proposed in Ref. \cite{b5c} for finite-size systems   can be  applied also   in  higher dimensions (in our case, due to the periodic boundary conditions,  the final mass is $m_f=0.01$ and thus the maximum quasiparticle velocity is almost one, the speed of light).       
\section{Universe expansion}
\label{s4}
In this section we study  the scalar field $\Phi$  in the curved spacetime. More  precisely, we consider   the FLRW metric 
\be
\label{u1}
ds^2=dt^2 -a^2(t)d\bx ^2=a^2(\eta)(d\eta^2-d\bx^2),
\ee
in the cosmic time   $t$; or  alternatively, in  the conformal time $\eta$,  $dt=a(\eta)d\eta$.
\par  
Let us start with  the $(1+1)$-dimensional   case and  conformal  time.     Then, following the standard discretization procedure   (with lattice spacing equals one)  we arrive at the Hamiltonian   
\be
\label{u2}
H(\eta)=\frac 12 \sum_j\left(\pi_j^2+(\phi_j-\phi_{j+1})^2+m^2(\eta)a^2(\eta)\phi_j^2\right).
\ee 
Let us note that the Hamiltonian \eqref{u2} under the Dirichlet boundary condition  can be written in the form \eqref{hd}  with $\Lambda(\eta)$ given by eq. \eqref{a1} with $M^2=m^2(\eta)a^2(\eta)$. Thus the eigenvalues of $\Lambda(\eta)$ read 
\be
\lambda_j(\eta)=\lambda_j^0+m^2(\eta)a^2(\eta),
\ee  
where $\lambda^0_j$ are the  (constant) eigenvalues of  the matrix \eqref{a1} with $M=0$, i.e. $\lambda_j^0=4\sin^2(j\pi/N)$, $j=1,\ldots,N$. 
\par 
Alternatively, for    the cosmic time   we  get 
\be 
\hat H(t)=\frac 12 \sum_j\left(\hat \pi_j^2+\frac{(\hat\phi_j-\hat\phi_ {j+1})^2}{a^2(t)}+\hat \Omega(t)\hat \phi_j^2\right),
\ee
where $\hat \Omega(t)=m^2(t)+\dot a^2/4a^2-\overset{..}{a}/2a$.  
The equivalence of the descriptions in both times  can be confirmed by the following  canonical time-dependent    transformation    
\be
\label{u3} 
\phi_j= \hat  \phi_j/\sqrt a, \quad p_j=\sqrt{a}\hat p_j-\dot a  \hat\phi_j/(2\sqrt a)\ .
\ee 
In fact,  we have the  identity 
\be
\label{u4}
H(\eta(t))\frac{d\eta}{dt}+\frac{\partial F}{\partial t}=\hat H(t),
\ee
where $F$ is the  generating function  of the transformation \eqref{u3}, i.e.  
\be
F(\phi_1,\ldots,\phi_N,\hat  p_1,\ldots,\hat p_N,t)=\sum_j(\sqrt{a} \phi_j\hat p_j-\dot a \phi_j\phi_j/4).
\ee
The above   transformation implies that both descriptions  are equivalent;    in consequence the symplectic covariance argument, see Ref. \cite{sitter},    implies that  the entropy and      capacity of entanglement    are equivalent for both realizations. Below,    we will show this  explicitly by  considering  the evolution of the initial ground state, see Sec. \ref{s1}.  Namely,  let us take the functions $b_j(\eta)$ defining  the evolution  of the state in the  conformal time, i.e.   satisfying the Ermakov equation \eqref{ee9} with $\lambda_j(\eta)=m^2(\eta)a^2(\eta)+\lambda_j^0$ . Then, we can readily check that  the functions 
\be
\hat  b_j(t)=b_j(\eta(t))\sqrt{a(t)},
\ee 
satisfy the Ermakov equation   in  the cosmic time (with $\hat\lambda_j(t)=\hat \Omega(t)+\lambda^0_j/a^2(t)$). This together with the results  presented in Sec. \ref{s1}  yield  $\hat{\tilde  B} =(\tilde B+\dot aI/4)/a$,  thus $\hat{ \tilde B}_2=\tilde B_2/a$; on the other hand, we have $\hat \Omega=\Omega/a$. In consequence,  by virtue of eq. \eqref{ee20a}   the eigenvalues  of the  matrix \eqref{ee21} (and thus \eqref{ee21a}) do not change and consequently the R\'enyi entropies as well (we have to make only the  substation  $S(\eta(t))=\hat S(t)$). In particular, when $m=0$ then  $b_j=1$ and thus $S(\eta)=const$. On the other hand, we have  $\hat b_j=\sqrt{a}$ and the matrix $\hat B$ is diagonal;  in consequence, $\hat B_2=0$  and then the matrix $\hat \Upsilon$ and $\hat \Delta$ contain the time dependency through  the same common factor   $1/a$;  consequently, $\hat \Delta$ is time-independent and  $\hat S$  and $\hat C$ are  also  constant. 
\par  
A similar situation holds   in  $1+3$ dimensions. Namely,  for the conformal time the discretization procedure (enhanced by  the simple canonical transformation i.e.  the scaling of momenta by $a$ and coordinates by $1/a$) yields that the  Hamiltonian $H$  is the sum $H(\eta)=\sum_{lm}H_{lm}(\eta)$ where
\be
\label{hsc}
H_{lm}(\eta)=\frac 1 2 \sum_j\left(\pi_{lm,j}^2+(j+\frac 12)^2\left(\frac{\phi_{lm,j}}{j}-\frac{\phi_{lm,j+1}}{j+1}\right)^2+\frac{l(l+1)}{j^2}\phi^2_{lm,j}+\Omega(\eta)\phi_{lm,j}^2\right),
\ee 
with $\Omega(\eta)=m^2(\eta)a^2(\eta)-\frac{a''(\eta)}{a(\eta)}$.
Thus, the Hamiltonian $H_{lm}(\eta)$ takes the form \eqref{hd} with $\Lambda$ given by eq. \eqref{a3} with $M^2=\Omega(\eta)$. In consequence,  the eigenvalues are  of the form $\lambda_j=\lambda_j^0+\Omega(\eta)$ where $\lambda_j^0$ are  the (constant) eigenvalues of the matrix \eqref{a3} with $M=0$.  
On the other hand, performing the canonical transformation \eqref{u3} we obtain the Hamiltonian in the cosmic time  $\hat H(t)=\sum_{lm}\hat H_{lm}(t)$.
\be
\hat H_{lm}(t)=\frac 1 2 \sum_j\left(\hat \pi_{lm,j}^2+\frac{(j+\frac 12)^2}{a^2(t)}\left(\frac{\hat \phi_{lm,j}}{j}-\frac{\hat \phi_{lm,j+1}}{j+1}\right)^2+\frac{l(l+1)}{j^2a^2(t)}\hat \phi^2_{lm,j}+\hat \Omega(t))\hat \phi_{lm,j}^2\right),
\ee 
where $\hat \Omega(t)=m^2(t) -\frac{3\dot a^2(t)}{4a^2(t)}-\frac{3\overset{..}{a}(t)}{2a(t)}$.
As above the dynamics of the initial  state    governed by  both the  Hamiltonians is  determined by the functions  $b_j(\eta)$ and $\hat b_j(t)$ respectively, which    satisfy the  suitable  Ermakov equations (with the  frequencies $\lambda_j$ and $\hat \lambda_j$, containing $ \Omega$ and $\hat\Omega$,  respectively). Now, by straightforward calculations we  check that  the   relation \eqref{u4} holds. In consequence, the suitable entropies coincide in  both pictures.  
\subsection{De Sitter space}
\begin{figure}[!ht]
	\begin{center}
		\includegraphics[width=0.99\columnwidth]{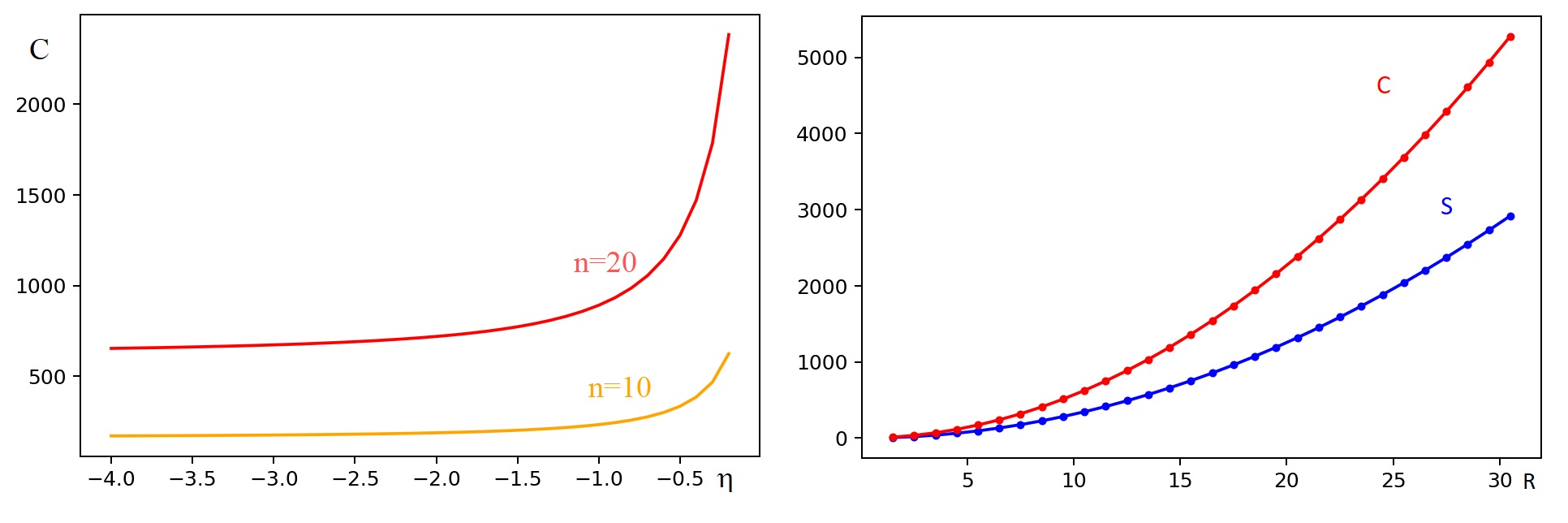}
	\end{center}
	\caption{\small{ $(1+3)$-dimensional dS space,   $N=60$, $m=0$, $H=1$. The left panel: capacity $n=10$ (yellow), $n=20$ (red). The  right panel:  the time-fixed slice $\eta=-0.2$ (together with the   parabolic fitting) -   entropy (blue)  and capacity (red).
			\label{ds1}}
	}
\end{figure}

\begin{figure}[!ht]
	\begin{center}
		\includegraphics[width=0.99\columnwidth]{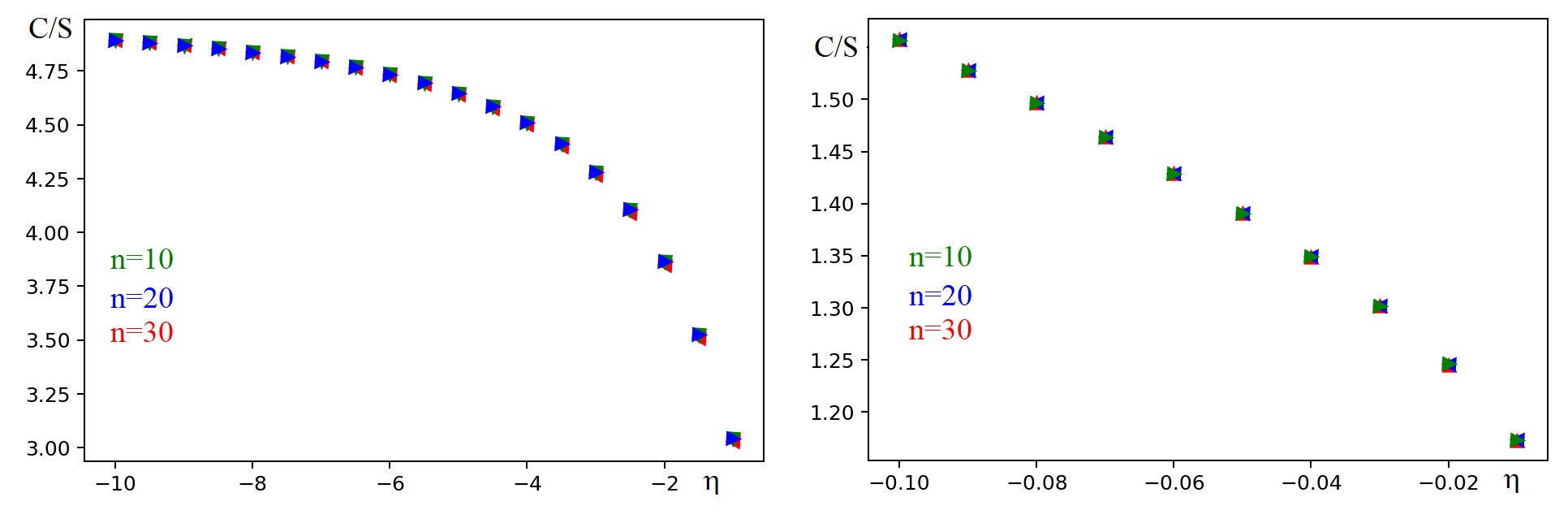}
	\end{center}
	\caption{\small{$(1+3)$-dimensional dS space. $N=60$, $m=0$, $H=1$.   The ratio of the capacity and entropy for $n=10$, (green) $n=20$ (blue), and $n=30$ (red). 
			\label{ds2}}
	}
\end{figure}

For  dS space we have $a(t)=e^{Ht}$, equivalently $a(\eta)=(-1)/H\eta$ for  $\eta<0$.  The dynamics of the entanglement entropy for  dS space  (and the standard vacuum states) was studied in   Ref. \cite{sitter2}  and more  recently in Ref. \cite{sitter0}   for  a field  at two distinct spatial locations \cite{sitter0b}, as well as  in the lattice approach  in Refs. \cite{sitter,sitter3,sitter4}.  In particular, it has been argued therein  that at the leading order the area law holds  with respect to the proper area of the surface; for  the  comoving coordinates the suitable $\eta$-dependence arises.
In what follows we  assume the  Bunch-Davies (BD) vacuum,  then the solutions of the Ermakov equations,  i.e. $b$'s functions, tend to one while  their derivatives tend  to zero for $\eta\rightarrow-\infty$.  In consequence, they read
\be
\label{es1}
b_j^2(\eta)=-\frac\pi 2 \sqrt{\lambda_j^0}\eta\left (J_\nu^2(-\eta\sqrt{\lambda_j^0})+Y^2_\nu(-\eta\sqrt{\lambda_j^0})\right),
\ee    
where, $J_\nu$ and $Y_\nu$ are the Bessel  functions, while, in $1+3$ dimensions, $\lambda^0_j$  are the eigenvalues of the matrix \eqref{a3} with $M=0$, while $\nu=\sqrt{9-4m^2/H^2}/2$. The  typical dynamics of the capacity   is presented in the left panel of Fig. \ref{ds1} (for $n=20$ and $n=30$)\footnote{To make contact with  our previous considerations,     we follow the   regularization procedure of counting $l$  from previous sections. However,  a different  regularization can also  be used, see Ref. \cite{cosmo},   then  subhorizon modes are excluded and $l$ is truncated earlier.}. Similarly to the entropy, the  capacity increases when $\eta$ approaches to zero; moreover, it increases with $n$ (radius).  To analyse the latter issue and  the  area  law, we  plot time slices  with respect to $n$. Then, we observe in   the right panel of Fig. \ref{ds1}, that   the area law holds also  for the capacity. Namely, for a fixed time (even  small  $\eta=-0.2$)  the  values   fit very well into the parabola. 
In view of this  the  ratio   of the capacity and entropy (at the leading order) does depend on the radius.  In  previous sections  for  massless field in  the $(1+3)$-dimensional  Minkowski  spacetime  we obtained  that this ratio  is equal to  $5.2$; now, let us analyse this problem for  dS space. To this end we plot the ratio $C/S$ for several values of $n$, see Fig. \ref{ds2}.
First, we  observer that the ratio indeed  does not depend on $n$ ($n=10,20,30$ coincide); moreover, it is constant  and approximately equals five for  initial  times (this is in agreement  with the   considerations presented in Sec. \ref{s2} and  the definition of the  BD  state).  
However,  as $\eta$  approaches to zero  $C/S$  is  decreasing. The numerical results, see the right plot in  Fig. \ref{ds2},  yields that    this ratio tends to one in the  limit $\eta\rightarrow 0^-$ : 
\be
\frac C S \simeq 5, \  \textrm{ for }  \eta \rightarrow -\infty , \quad \frac C S \simeq  1,\  \textrm{ for }  \eta \rightarrow 0^- \ . 
\ee 
According to the results obtained in Ref.  \cite{boer},  the above   observation may  suggest  that    the scalar field theory  in dS space at late times  is      related to    conformal field theories (despite the fact that in full dS  space we do not have the conformal covariance, i.e. in general $m^2\neq 2H^2$).     This idea    fits into the recent  investigations of  dS/CFT correspondence  \cite{latetime1,latetime2,latetime3}, where  the behaviour of the scalar fields in dS space  at  late times (i.e. $\eta\rightarrow 0^-$)  is associated   with      certain  (Euclidean) conformal field theories (more precisely,  generating functional for correlation functions of the dual conformal  theories).  Although the description of this correspondence is not complete currently,  following  the    AdS/CFT correspondence,    a  further analysis  of  the entanglement structure   at late times in dS space  would be   valuable.   
\subsection{Radiation-dominated era} 
\begin{figure}[!ht]
	\begin{center}
		\includegraphics[width=0.99\columnwidth]{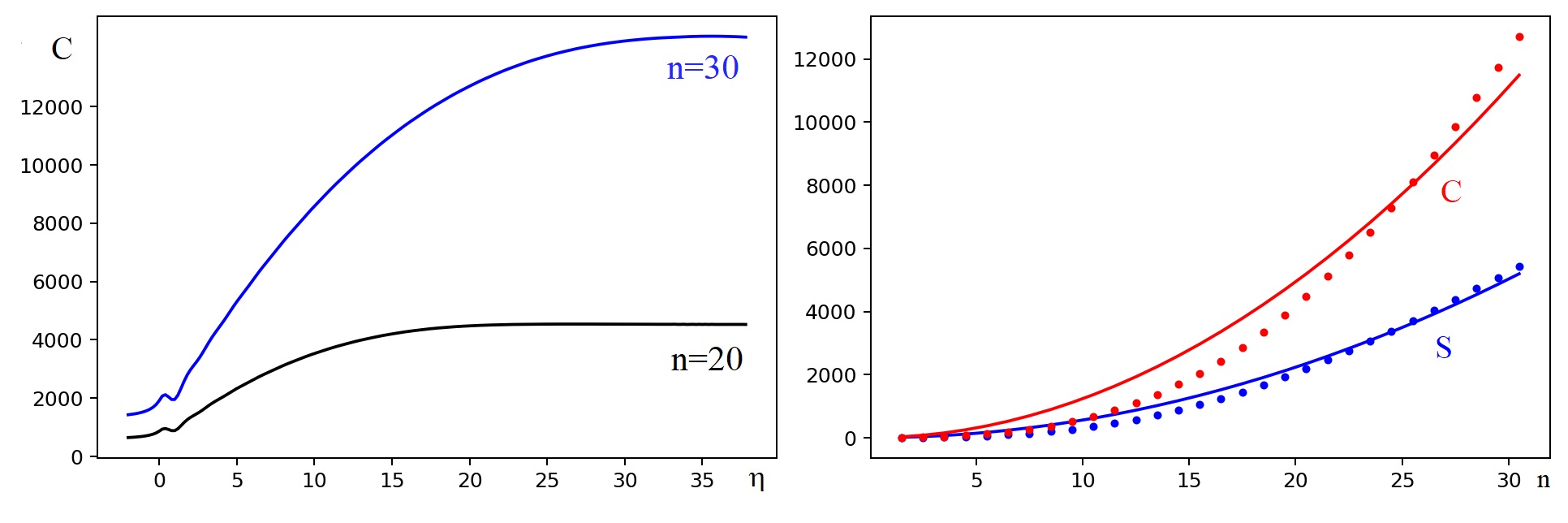}
	\end{center}
	\caption{\small{ The  transition from  $(1+3)$-dimensional dS space to the radiation-dominated era (at $\eta=-1$); $N=60$, $m=0$, $H=1$. The left panel: evolution of the capacity $n=20$ (black), $n=30$ (blue). The  right panel:  time-fixed slice for $\eta=50>0$ and the parabolic fitting -   entropy (blue)  and capacity (red).
			\label{rd1}}
	}
\end{figure}

\begin{figure}[!ht]
	\begin{center}
		\includegraphics[width=0.99\columnwidth]{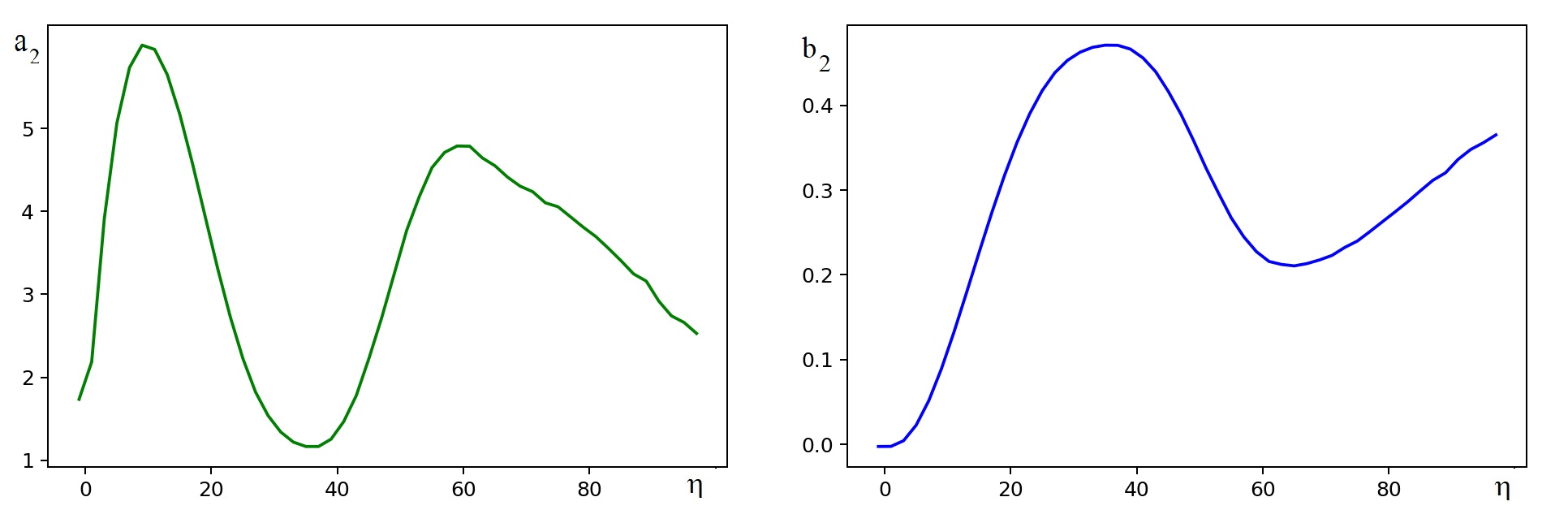}
	\end{center}
	\caption{\small{ The  transition from  $(1+3)$-dimensional dS space to the radiation-dominated era (at $\eta=-1$); $N=60$, $m=0$, $H=1$. The  temporal evolution  of the capacity. The left plot: the square factor $a_2$. The right plot: the cubic factor $b_2$, cf. eq. \eqref{dec}. 
			\label{rd2}}
	}
\end{figure}

Now, we let us consider  other FLRW metrics. Namely, we will study  the   transition from   dS space to  the radiation-dominated era; the latter will be  modeled by a  metric with linear function $a(\eta)\sim \eta$ (by similar considerations we can add the  era of matter domination, $a(\eta)$ is a quadratic function). In such a case, $a''(\eta)=0$ and thus for the massless field $\Omega=0$; then, in turn, the function $b_j$ can be readily found.   It turns out that, after  the transition  the area law breaks  and a volume term develops giving  contribution to the entropy at late times \cite{cosmo,cosmo2}.  A similar situation appears for the capacity.  Namely, assuming the transition from  dS to the RD era takes  place at $\eta=-1$, we observe  that the monotonic growth is broken and  the quasi-periodicity  appears during the evolution, see the  left panel in  Fig. \ref{rd1} for the initial times. Moreover, after transition   the  area  law does not hold  (the contribution from the volume term appears), see dots in the right panel in  Fig.  \ref{rd1} where the entropy and capacity for $\eta=50$ are depicted together with suitable parabolas.   
To analyse this situation in more detail  we split the dynamics of the capacity into  two  parts:   $a_2$ and $b_2$, related  to the quadratic and cubic parts, see eq. \eqref{dec}.  The temporal evolution of these coefficients are presented in     Fig. \ref{rd2}. We observe that   at the beginning the  area law holds with good approximation; however, for further times the cubic term develops, and the situation repeats quasi-periodically; finally,  the oscillations of the cubic part decay with time and asymptotically settle to a constant value (in our case $b_2\simeq 0.3$).     
\subsection{Quenches in de Sitter space}
Now,  we will follow  Sec. \ref{s2}  and  consider  quench phenomena in  dS space. 
This is interesting     due to the fact that   rapid phenomena   during the  early expansion of the universe  (e.g.   inflation,  reheating,  particles production)  should be related to     the  non-equilibrium processes, see e.g. \cite{cosmo3,cosmo4,cosmo5} and references therein.    To get some insight into these issues, in particular  thermalisation  (equilibration),        quenches in  closed systems are frequently considered.    In  view of this,   we will study  the evolution of entanglement entropy and capacity by means of an abrupt quench in dS space; with the special attention put  on the    transition from the  area law   to volume law (the latter is characteristic  for  thermal entropy). 
\par 
To this end, let us recall  that  the mass quench in the  FLRW spacetime is  described by the Hamiltonian \eqref{hsc} with effective mass $\Omega(\eta)=m^2(\eta)a^2(\eta)-a''(\eta)/a(\eta)$. Thus,  the mass quench  in the FLRW  space can be recast as the    mass  quench (relatively to effective mass $\Omega$) in the  ordinary Minkowski spacetime (if we add conformal coupling term  $\xi R$ to the action, then  we get   a more direct expression  $\Omega(\eta)=m^2(\eta)a^2(\eta)$).  Consequently,  in this picture,   the abrupt quench in the FLRW space (in particular for dS space)  does not give an instantaneous  cooling in the Minkowski spacetime needed for  the quasiparticles model. 
Moreover,  as mentioned  in Sec. \ref{s2},    for the Minkowski spacetime  in  three spatial dimensions we encounter some additional  problems; for  dS space  the situation becomes more complicated, i.e. the entropy contains  additional terms related  to the Hubble constant (in particular the long range contribution  to the entanglement entropy). 
In view of this, to analyse the entanglement dynamics,   we follow the lattice approach  presented in previous sections.  To this end we need the functions $b_j$ describing the  dynamics of the reduced density. 
\par
\begin{figure}[!ht]
	\begin{center}
		\includegraphics[width=0.99\columnwidth]{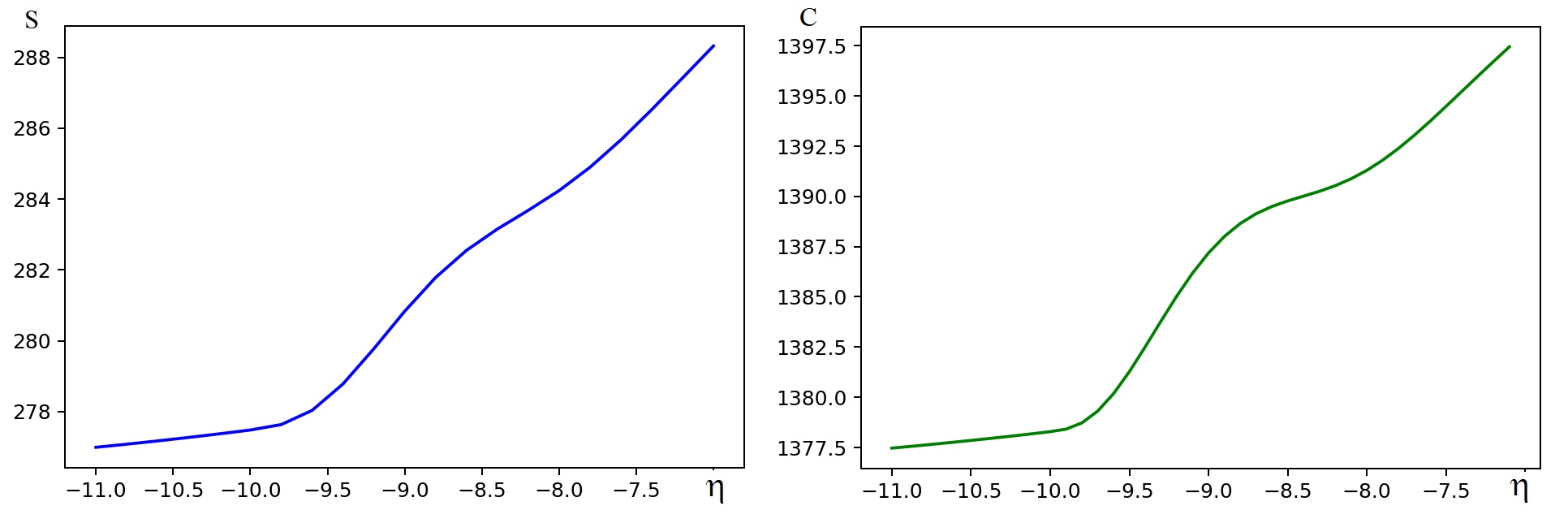}
	\end{center}
	\caption{\small{The quench in  $(1+3)$-dimensional dS space; $N=60$, $H=1$, the mas quench at $\eta_0=-10$,  the initial mass $m_i=\sqrt{5}/2$ and   $n=30$. The left plot: entropy, the right plot: capacity. 
			\label{qs}}
	}
\end{figure}
\begin{figure}[!ht]
	\begin{center}
		\includegraphics[width=0.99\columnwidth]{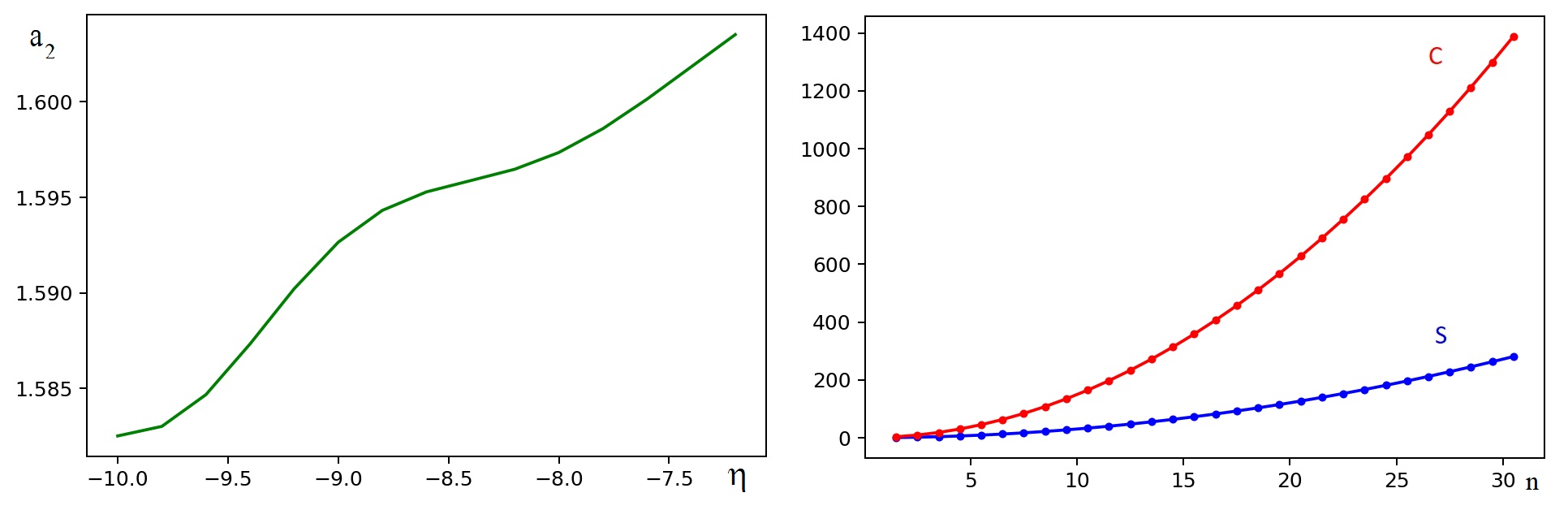}
	\end{center}
	\caption{\small{The quench in  $(1+3)$-dimensional dS space; $N=60$, $H=1$, the mass quench at $\eta_0=-10$,  the initial mass $m_0=\sqrt{5}/2$. The left plot:  the quadratic  factor $a_2$ for the capacity. The right panel: entropy (blue)  and capacity (red) -- the time-fixed  slice (at $\eta=-9>\eta_0$) after quench (together with the  parabolic fitting).   
			\label{qs2}}
	}
\end{figure}

More specifically, let us  assume that we start with   a massive field $m_i$ and next   there is an  abrupt   change  of the mass parameter to zero value (the massless field). Then,  the frequencies appearing in the Ermakov equation   are give by the formula 
\be
\label{dsq}
\lambda_j(\eta)=\lambda_j^0+\left\{
\begin{array}{c}
	(\frac{m_i^2}{H^2}-2)\frac {1}{\eta^2}, \quad \eta<\eta_0,\\
	-\frac{2}{\eta^2} \quad \eta \in [\eta_0,0),
\end{array}
\right. 
\ee
where $\eta_0<0$ is a fixed point. 
Assuming  the BD  vacuum state  the functions $b_j$ are described by   eq. \eqref{es1}  for $\eta<\eta_0$ .  It remains to  find $b_j$ for $\eta\in [\eta_0,0)$ in such  a way that  they  as well their  derivatives are continuous  at the point $\eta_0$. This can be done by the straightforward but rather  tedious computations. The final result reads
\be
\label{dsq2}
b_j(\eta)=\sqrt{x_j^2(\eta)+y_j^2(\eta)/A^2} , \quad \eta\in[\eta_0,0),
\ee      
where the functions $x_j(\eta),y_j(\eta)$ and the constant $A$ are given in  Appendix \ref{s7}. Now, we are in the position to analyse the dynamics  of the entropy and  capacity. Of course, for $\eta<\eta_0$,  we have the  monotonically increasing growth of the entropy and capacity.  Since the  frequencies $\lambda_j$ in both cases contain the factor $1/\eta^2$ we expect also a similar behaviour for $\eta\rightarrow 0$. However, for the intermediate times (related to the initial mass)  this situation may  change. Indeed,   we observe  in Fig. \ref{qs} that  after the change of mass (here, from $m_i=\sqrt{5}/2$ to zero)  at $\eta_0=-10$,      there is a period   of time resembling quasi-oscillatory behaviour in the Minkowski spacetime, and next (for $\eta>-7$) again both quantities  uniformly increase.   
However, it seems that  there is a  one difference to the Minkowski case; namely, for these  intermediate times the area law is preserved with good approximation. Indeed, for both the  entropy and capacity the coefficients corresponding to the cubic terms are very small compared to the  quadratic terms  (i.e.  $b_i/a_i\simeq 0.001$ for $i=1,2$); in consequence, the   leading  contribution  comes from the quadratic term,  see Fig. \ref{qs2}, where  the evolution of the quadratic factor $a_2$  for the  capacity is presented  (cf. the right plot in Fig. \ref{qs}) as  well as a time fixed slice together  with the parabola approximation for $\eta=-9$  (an intermediate time).   Consequently, the area law holds for all times with good accuracy.  
This situation is different from that obtained in   Sec. \ref{s2}, where  the volume term  after quench was significant.   Similarly to  the free theory   in the Minkowski spacetime    (in general, for  the integrable models),  we expect that   the  thermalisation takes place  in  a more general sense, i.e.  the steady state is    described by  a generalized  Gibbs ensemble (due to  the constrains related to the conserved quantities), see e.g. \cite{gge,gge2} and references therein.     However, in dS space the  absence of volume term implies  further complications;      the injection of energy does not essentially  change  the entropy, in  other  words, entropy   attains the   maximum value; consequently,  it  is  related to a uniform  probability distribution. This, in turn, suggest that   thermalisation takes place for  (large) infinite temperature. Although, such a   reasoning is quite superficial,  it has some points in common with    recent ideas concerning  thermodynamic of  dS space presented in  Ref. \cite{dst}. To get more deeper insight into the above issue and  to better  understand non-equilibrium processes in dS space further studies (including continuous quenches, interacting fields and the symmetry breaking mechanisms) are needed.       
\section{Conclusions}
\label{s5} 
In this work, we have studied   some  aspects of entanglement  in quantum  field theory; in particular, the ones related to the area law.  To get   more insight into these issues, besides   entropy, we use  the notion of the capacity  of entanglement which     gained recently some attention and can be treated as a measure of  entropy fluctuations.  Both the quantities  together  can provide more information on the entanglement structure  and   can be helpful in finding some universal properties  of the theory.  In our  investigations, we have considered   physically more interesting, but less commonly studied, higher dimensional spacetimes  including    some curved backgrounds (relevant for cosmology). Moreover,  the special attention has been put on   quench  phenomena,   which are  useful  in various physics contexts (such as  thermalisation processes, phase transitions and the physics of  early universe).  In the latter  case,  the  time-dependent mass parameter has been  considered and the   numerical results were compared with  the dynamics resulting  from  theoretical models. Let us now  summarize our results. 
\par 
First, we showed that for the fields with constant  mass, the capacity, like entropy, exhibits the area law (at the leading order).  We  observe this for two kinds of geometries of the   entangling surface in the Minkowski spacetime: spheres and strips. This observation  implies that the ratio of both quantities  does not depend on the area, and more importantly,  this ratio   takes the  same values  for both kinds of geometries. 
Next, we turned our attention to quenches. First, we analysed the dynamics of capacity and showed that  after some initial time the area law is broken and the volume term comes into play;  for sufficiently   strong abrupt quenches this term can be crucial. Moreover,  for the  strip geometry the initial  growth  is linear    and does not  depend on  the  width of the strip (thus the area law holds). To get some insight into this issue we compared these results with theoretical predictions resulting  from the quasiparticles model;  in  $1+2$ dimensions we  got  good agreement  for both geometries (including the revivals times  for the entanglement dynamics). 
\par  
In the second part we considered  the above issues  in  curved spacetimes. We started by  showing  explicitly   that for fields in  the  FLRW  space the   description of the entropy  (capacity)  in the cosmic and conformal times are equivalent. Next, we specialized the metric taking  dS space, and then   the metric   modeling   transition to the radiation-dominated era.  For  the BD state of dS space, similarly to the  entropy,  the area scaling  holds also  for the capacity.  Thus the ratio $C/S$   does not depend on the radius and  it  tends to one as conformal time approaches zero; the latter behaviour is interesting due to the recent investigations of dS/CFT correspondence.  Moreover, we showed that after the transition to the radiation-dominated era the volume term develops in the dynamics of the capacity. 
Finally, we analysed the evolution of the entropy and capacity during the  abrupt quench in dS space. First,  we found the functions describing the evolution of the  state in such a scenario. Next,  using these  results, we   showed that after quench   the area law survives  with good accuracy. 
\par   Of course, the above results do not exhaust the subject. In this context  it  would be interesting to consider continuous or  
multiple quenches \cite{b12d}, higher derivative theories \cite{b20a} or different  vacuum states  in dS space  \cite{b20b}.   The notion of the modular entropy \cite{b20c} and    boundary  quenches  can be also examined \cite{cotler}. Finally, following Refs. \cite{b20d,b20e} the consequences of non-minimal  coupling   terms  and regular black holes   are also worth of study.

\acknowledgments
The  author would like to thank   Mohammad   Reza Mohammadi Mozaffar   for valuable discussion and references  as well as  Piotr  Kosi\'nski for  useful  remarks. 
Special thanks are due to anonymous referee for valuable comments and suggestions. 

\appendix
\section{Discretization procedure }
\label{s6}
In this appendix,  we briefly recall the form of the discretized  Hamiltonians and the corresponding entropies for  two geometries of the entanglement surface  in the Minkowski spacetime, see e.g. Refs.  \cite{cotler,b13a,b13b,b13c}.  These facts  turn out to be useful  for the  FLRW metrics presented in Sec. \ref{s4}. 
\par 
Let us   consider the scalar field with mass $M$. In 1+1 dimensions  (and the Dirichlet boundary conditions)  the matrix $\Lambda$  in eq. \eqref{hd}  is of the form  
\be
\label{a1}
\Lambda_{jj}=2+M^2, \quad \Lambda_{j,j+1}=\Lambda_{j+1,j}=-1.
\ee
In 1+2 dimensions the  discretized Hamiltonian is the sum $H=\sum_{l=-\infty}^{\infty} H^l$ where $H^l$ are  of the form \eqref{hd} with the following $\Lambda^l$  
\be
\label{a2}
\Lambda^l_{11}= \frac 3 2+l^2+M^2, \quad \Lambda^l_{jj}= 2+\frac{l^2}{j^2}+M^2 , \quad \Lambda^l_{j,j+1}=\Lambda^l_{j+1,j}=-\frac{(j+ 1/2)}{\sqrt{j(j+1)}} .
\ee
In view of this the R\'enyi  entropy for the Gaussian state   is the sum of the $l$ components  $R_\alpha=R_\alpha^0 +2\sum_{l=1}^\infty R_\alpha^l$; in consequence, the same holds for the entropy and capacity. 
\par 
For  $1+3$ dimensions and the spherical geometry we  have $H=\sum_{l=0}^{\infty}\sum_{m=-l}^l H^{lm}$ where $H^{lm}$ is  described by \eqref{hd} with  
\be
\label{a3}
\Lambda^{lm}_{11}=\frac 9 4+l(l+1)+M^2,\quad \Lambda^{lm}_{jj}=2+\frac{1}{(2j^2)}+\frac{l(l+1)}{j^2}+M^2, \quad \Lambda^{lm}_{j,j+1}=\Lambda^{lm}_{j+1,j}=-\frac{(j+1/2)^2}{j(j+1)}.
\ee
Thus $R_\alpha=\sum_{l=0}^\infty (2l+1)R_\alpha^l$  and analogously for the entanglement entropy and capacity. 
\par
For the strip geometry in  the $(1+d)$-dimensional Minkowski spacetime  we  trace over  a $d$-dimensional slab of width $2R$ and the cross-sectional area $A_\perp$. Then,  see \cite{cotler},  the  Hamiltonian factorizes 
\be
H=\frac{A_\perp}{ (2\pi)^{d-1}}\int d^{d-1}k_\perp \tilde H(q_{k_\perp},p_{k_\perp}, M^2+ k_\perp^2),
\ee  where $\tilde H$ is the Hamiltonian of a massive (with the mass parameter $M^2+k_\perp^2$)  field in $1+1$ dimensions;  $k_\perp$ denotes momentum  in  the perpendicular direction. The  modes $k_\perp $ decouple and, consequently, the R\'enyi entropy can be  reduced to the integral over  one-dimensional counterparts. This together with  formula \eqref{ec}   give the  capacity described by eq.   \eqref{capnu}. 
\section{Quench of  the Bunch-Davies vacuum }
\label{s7}  
In this appendix we compute the  solutions of the Ermakov equations describing the  quench of the BD state  in $(1+3)$-dimensional  dS space. Namely, let us assume that  mass changes at $\eta=\eta_0<0$ from $m_i$ to   zero, and for $\eta<\eta_0$ we have the BD  vacuum state.   Then   $\lambda_j(\eta)$ are given by eq. \eqref{dsq} where    $\lambda^0_j$  are the eigenvalues of the matrix \eqref{a3} with $M=0$. Moreover, for $\eta\leq\eta_0$ the solutions of the  Ermakov equations \eqref{ee9} are given by formula \eqref{es1}. Now, we will find $b$'s   after quench (demanding that their derivatives are continuous at $\eta_0$). After straightforward  but tedious computations, we get that the functions $b_j(\eta)$ for $\eta\in[\eta_0,0)$  are given by eq. \eqref{dsq2}  with 
\be  
x_j(\eta)=C_j \left (\frac{\sin( \sqrt{\lambda_j^0 } \eta )}{ \sqrt{\lambda_j^0 } \eta }-\cos( \sqrt{\lambda_j^0 } \eta )\right)+D_j \left(\sin( \sqrt{\lambda_j^0 } \eta )+\frac{\cos( \sqrt{\lambda_j^0 } \eta )}{ \sqrt{\lambda_j^0 } \eta }  \right) ,
\ee
\be
\begin{split}
	y_j(\eta)=&\left(\sin( \sqrt{\lambda_j^0 } \eta_0)+\frac{\cos( \sqrt{\lambda_j^0 } \eta_0)}{ \sqrt{\lambda_j^0 } \eta_0}\right)\cdot \left(\cos( \sqrt{\lambda_j^0 } \eta)-\frac{\sin( \sqrt{\lambda_j^0 } \eta)}{ \sqrt{\lambda_j^0 } \eta }\right)+\\
	&\left(\frac{\sin( \sqrt{\lambda_j^0 } \eta_0)}{ \sqrt{\lambda_j^0 } \eta_0}-\cos( \sqrt{\lambda_j^0 } \eta_0)\right) \cdot\left(\sin( \sqrt{\lambda_j^0 } \eta)+\frac{\cos( \sqrt{\lambda_j^0 } \eta)}{ \sqrt{\lambda_j^0 } \eta}\right),
\end{split}
\ee
where 
\be
\begin{split}
	C_j=&\frac{A_j}{\lambda_j^0}\left (\frac{\cos(\sqrt{\lambda_j^0} \eta_0)}{\eta_0^2 }+\frac{\sqrt{\lambda_j^0}}{\eta_0} \sin(\sqrt{\lambda_j^0} \eta_0)-\lambda_j^0 \cos(\sqrt{\lambda_j^0} \eta_0)\right) +\\
	&       \frac{B_j}{\lambda_j^0}\left (\sqrt{\lambda_j^0} \sin(\sqrt{\lambda_j^0} \eta_0)+\frac{\cos(\sqrt{\lambda_j^0} \eta_0)}{\eta_0}\right), 
\end{split} 
\ee
\be
\begin{split}
	D_j=&\frac{A_j}{\lambda_j^0} \left(\lambda_j^0 \sin(\sqrt{\lambda_j^0} \eta_0)-\frac{\sin(\sqrt{\lambda_j^0} \eta_0)}{\eta_0^2}+\frac{\sqrt{\lambda_j^0}}{\eta_0} \cos(\sqrt{\lambda_j^0} \eta_0)\right)+\\
	& \frac{B_j}{\lambda_j^0}\left (\sqrt{\lambda_j^0} \cos(\sqrt{\lambda_j^0} \eta_0)-\frac{\sin(\sqrt{\lambda_j^0} \eta_0)}{\eta_0}\right) ,
\end{split}
\ee
and $b_j(\eta_0)=A_j$, $\frac{db_j}{d\eta}(\eta_0)=B_j$ are the values for the BD state at $\eta=\eta_0$ (see eq.  \eqref{es1}).  

\end{document}